\documentstyle[12pt,epsf,axodraw]{article}
\begin{document}
\pagestyle{plain} \setcounter{page}{1} \baselineskip=0.3in
\begin{titlepage}
\begin{flushright}
PKU-TH-99-68\\ NUHEP-TH-99-28
\end{flushright}
\vspace{.5cm}

\begin{center}
{\Large Supersymmetric Electroweak Corrections to Charged Higgs
\\ Boson Production in Association with a Top Quark \\ at Hadron
Colliders }

\vspace{.2in}
   Li Gang Jin $^a$, Chong Sheng Li $^{a}$, Robert J. Oakes $^b$ and
         Shou Hua Zhu $^{c,d}$  \\
\vspace{.2in}

$^a$ Department of Physics, Peking University, Beijing 100871,
China \\ $^b$ Department of Physics and Astronomy, Northwestern
University,\\ Evanston, IL 60208-3112, USA\\ $^c$ CCAST(World
Laboratory), Beijing 100080, China\\ $^d$ Institute of Theoretical
Physics, Academia Sinica, Beijing 100080, China \\
\end{center}
\vspace{.2in}
\begin{footnotesize}
\begin{center}\begin{minipage}{5in}
\baselineskip=0.25in
\begin{center} ABSTRACT \end{center}

We calculate the $O(\alpha_{ew}m_{t(b)}^{2}/m_{W}^{2})$ and
$O(\alpha_{ew} m_{t(b)}^4/m_W^4)$ supersymmetric electroweak
corrections to the cross section for the charged Higgs boson
production in association with a top quark at the Tevatron and the
LHC. These corrections arise from the quantum effects which are
induced by potentially large Yukawa couplings from the Higgs
sector and the chargino-top(bottom)-sbottom(stop) couplings,
neutralino-top(bottom)-stop(sbottom) couplings and charged
Higgs-stop-sbottom couplings. They can decrease or increase the
cross section depending on $\tan\beta$ but are not very sensitive
to the mass of the charged Higgs boson for high $\tan\beta$. At
low $\tan\beta(=2)$ the corrections decrease the total cross
sections significantly, which exceed $-12\%$ for $m_{H^{\pm}}$
below $300GeV$ at both the Tevatron and the LHC, but for
$m_{H^{\pm}}>300GeV$ the corrections can become very small at the
LHC. For high $\tan\beta(=10,30)$ these corrections can decrease
or increase the total cross sections, and the magnitude of the
corrections are at most a few percent
at both the Tevatron and the LHC.

\end{minipage}\end{center}
\end{footnotesize}
\vfill


\end{titlepage}

\eject \baselineskip=0.3in
\begin{center} {\Large 1. Introduction}\end{center}

   There has been a great deal of interest in the charged Higgs bosons
appearing in the two-Higgs-doublet models(THDM)[1], particularly
the minimal supersymmetric standard model(MSSM)[2], which predicts
the existence of three neutral and two charged Higgs bosons $h, H,
A,$ and $H^{\pm}$. When the Higgs boson of the Standard Model(SM)
has a mass below 130-140 Gev and the h boson of the MSSM is in the
decoupling limit (which means that $H^{\pm}$ is too heavy anyway
to be possibly produced), the lightest neutral Higgs boson may be
difficult to distinguish from the neutral Higgs boson of the
standard model(SM). But charged Higgs bosons carry a distinctive
signature of the Higgs sector in the MSSM. Therefore, the search
for charged Higgs bosons is very important for probing the Higgs
sector of the MSSM and, therefore, will be one of the prime
objectives of the CERN Large Hadron Collider(LHC). At the LHC the
integrated luminosity is expected to reach $L=100 fb^{-1}$ per
year in the second phase. Recently, several studies of charged
Higgs boson production at hadron colliders have appeared in the
literature[3,4,5]. For a relatively light charged Higgs boson,
$m_{H^{\pm}}< m_t - m_b$, the dominate production processes at the
LHC are $gg\rightarrow t \bar t$ and $q\bar q\rightarrow t\bar t$
followed by the decay sequence $t\rightarrow bH^+\rightarrow b\tau
^+\nu_{\tau}$[6]. For a heavier charged Higgs boson the dominate
production process is $gb\rightarrow tH^-$[7,8,9]. Previous
studies showed that the search for heavy charged Higgs bosons with
$m_{H^{\pm}}>m_t + m_b$ at a hadron collider is seriously
complicated by QCD backgrounds due to processes such as
$gb\rightarrow t\bar tb, g\bar b\rightarrow t\bar t\bar b$, and
$gg\rightarrow t\bar tb\bar b$, as well as others process[8].
However, recent analyses[10,11] indicate that the decay mode
$H^+\rightarrow \tau ^+\nu$ provides an excellent signature for a
heavy charged Higgs boson in searches at the LHC. The discovery
region for $H^{\pm}$ is far greater than had been thought for a
large range of the $(m_{H^{\pm}}, \tan \beta)$ parameter space,
extending beyond $m_{H^{\pm}}\sim 1 TeV$ and down to at least
$\tan\beta \sim 3$, and potentially to $\tan\beta \sim 1.5$,
assuming the latest results for the SM parameters and parton
distribution functions as well as using kinematic selection
techniques and the tau polarization analysis[11]. Of course, it is
just a theoretical analysis and no experimental simulation has
been performed to make the statement very reliable so far.

   The one-loop radiative corrections to $H^-t$ associated
production have not been calculated, although this production
process has been studied extensively at tree-level[7,8,9]. In this
paper we present the calculations of the
$O(\alpha_{ew}m_{t(b)}^{2}/m_{W}^{2})$ supersymmetric(SUSY)
electroweak corrections to this associated $H^-t$ production
process at both the Fermilab Tevatron and the LHC in the MSSM.
These corrections arise from the quantum effects which are induced
by potentially large Yukawa couplings from the Higgs sector and
the chargino-top(bottom)-sbottom(stop) couplings, neutralino-
top(bottom)-stop(sbottom) couplings and charged Higgs-stop-sbottom
couplings which will contribute at the
$O(\alpha_{ew}m_{t(b)}^{4}/m_{W}^{4})$ to the self-energy of the
charged Higgs boson. In order to get a reliable estimate this
process has to be merged with the related gluon splitting
contribution $gg\rightarrow H^-t\bar{b}$. This leads to a
suppression by about $50\%$ at LO[12]. However, the complete
one-loop QCD corrections are probably more important, but not yet
available. \vspace{1cm}

\begin{center} {\Large 2. Calculations}\end{center}
\vspace{.3cm} The tree-level amplitude for $gb\rightarrow tH^-$ is
\begin{equation}
M_{0}=M_{0}^{(s)}+M_{0}^{(t)},
\end{equation}
where $M_{0}^{(s)}$ and $M_{0}^{(t)}$ represent the amplitudes
arising from diagrams in Fig.1$(a)$ and Fig.1$(b)$, respectively.
Explicitly,

\begin{eqnarray}
M_{0}^{(s)}&=&\frac{igg_{s}}{\sqrt{2}m_{W}(\hat{s}
-m_{b}^{2})}\overline{u}(p_{t})[2m_{t}\cot\beta p_{b}^{\mu}P_{L}
+2m_{b}\tan\beta p_{b}^{\mu}P_{R} -m_{t}\cot\beta
\gamma^{\mu}{\not{k}}P_{L} \nonumber \\ & &  -m_{b}\tan\beta
\gamma^{\mu}{\not{k}} P_{R}]u(p_{b})\varepsilon_{\mu}(k)
T_{ij}^{a},
\end{eqnarray}
and
\begin{eqnarray}
M_{0}^{(t)}&=& \frac{igg_{s}}{\sqrt{2}m_{W}(\hat{t}
-m_{t}^{2})}\overline{u}(p_{t})[2m_{t}\cot\beta p_{t}^{\mu}P_{L}
+2m_{b}\tan\beta p_{t}^{\mu}P_{R} -m_{t}\cot\beta
\gamma^{\mu}{\not{k}} P_{L} \nonumber \\ & &  -m_{b}\tan\beta
\gamma^{\mu}{\not{k}}P_{R}]u(p_{b})\varepsilon_{\mu}(k)T_{ij}^{a},
\end{eqnarray}
where $T^{a}$ are the $SU(3)$ color matrices and $\hat{s}$ and
$\hat{t}$ are the subprocess Mandelstam variables defined by $$
\hat{s}=(p_{b}+k)^2=(p_t+p_{H^-})^2,$$ and
$$\hat{t}=(p_t-k)^2=(p_{H^-}-p_b)^2. $$ Here the
Cabbibo-Kobayashi-Maskawa matrix element $V_{CKM}[bt]$ has been
taken to be unity.

The SUSY electroweak corrections of order
$O(\alpha_{ew}m_{t(b)}^{2}/m_{W}^{2})$ 
and $O(\alpha_{ew} m_{t(b)}^4/m_W^4)$ to the process
$gb\rightarrow H^-t$ arise from the Feynman diagrams shown in
Figs.1(c)-1(v) and Fig.2. We carried out the calculation in the
t'Hooft-Feynman gauge and used dimensional 
reduction, which preserves supersymmetry, for
regularization of the ultraviolet divergences
in the virtual loop
corrections using the on-mass-shell renormalization scheme[13], in
which the fine-structure constant $\alpha_{ew}$ and physical
masses are chosen to be the renormalized parameters, and finite
parts of the counterterms are fixed by the renormalization
conditions. The coupling constant $g$ is related to the input
parameters $e, m_W,$ and $m_Z$ by $g^2= e^2/s_w^2$ and
$s_w^2=1-m_w^2/m_Z^2$. The parameter $\beta$ in the MSSM we are
considering must also be renormalized. Following the analysis of
ref.[14], this renormalization constant was fixed by the
requirement that the on-mass-shell $H^{+}\bar l\nu_l$ coupling
remain the same form as in Eq.(2) of ref.[14] to all orders of
perturbation theory. Taking into account the
$O(\alpha_{ew}m_{t(b)}^{2}/m_{W}^{2})$ Yukawa corrections, the
renormalized amplitude for the process $gb\rightarrow tH^{-}$ can
be written as
\begin{eqnarray}
M_{ren}&=& M_{0}^{(s)} +M_{0}^{(t)} +\delta M^{V_{1}(s)} +\delta
M^{V_{1}(t)} +\delta M^{s(s)} +\delta M^{s(t)} +\delta
M^{V_{2}(s)} \nonumber \\ & & +\delta M^{V_{2}(t)} +\delta
M^{b(s)} +\delta M^{b(t)} \equiv M_{0}^{(s)} +M_{0}^{(t)}
+\sum_{l} \delta M^{l},
\end{eqnarray}
where $\delta M^{V_1(s)},\delta M^{V_1(t)}, \delta M^{s(s)},
\delta M^{s(t)},\delta M^{V_2(s)}, \delta M^{V_2(t)},\delta
M^{b(s)}$, and $\delta M^{b(t)}$ represent the corrections to the
tree diagrams arising, respectively, from the $gbb$ vertex diagram
Fig.1(c)-1(d), the $gtt$ vertex diagram Fig.1(f)-1(g), the bottom
quark self-energy diagram Fig.1(i), the top quark self-energy
diagram Fig.1(k), the $btH^-$ vertex diagrams Figs.1(m)-1(n) and
Figs.1(p)-1(q), including their corresponding counterterms
Fig.1(e), Fig.1(h), Fig.1(j), Fig.1(l), Fig.1(o), and Fig.1(r),
and the box diagrams Figs.1$(s)-1(v)$. $\sum_{l} \delta M^{l}$
then represents the sum of the contributions to the Yukawa
corrections from all the diagrams in Figs.1(c)-1(v). The explicit
form of $\delta M^{l}$ can be expressed as
\begin{eqnarray}\label{}
\delta M^{l}&=& -\frac{ig^{3}g_{s} T_{ij}^{a}}
{4\sqrt{2}\times16\pi^{2}m_{W}} C^{l}\overline{u}
(p_{t})\{f_{1}^{l} \gamma^{\mu}P_{L} +f_{2}^{l} \gamma^{\mu}P_{R}
+f_{3}^{l}p_{b}^{\mu}P_{L} +f_{4}^{l}p_{b}^{\mu}P_{R}
+f_{5}^{l}p_{t}^{\mu}P_{L} \nonumber \\ & &
+f_{6}^{l}p_{t}^{\mu}P_{R} +f_{7}^{l}\gamma^{\mu}{\not{k}}P_{L}
+f_{8}^{l} \gamma^{\mu}{\not{k}}P_{R}
+f_{9}^{l}p_{b}^{\mu}{\not{k}}P_{L}
+f_{10}^{l}p_{b}^{\mu}{\not{k}}P_{R}
+f_{11}^{l}p_{t}^{\mu}{\not{k}}P_{L} \nonumber \\ & &
+f_{12}^{l}p_{t}^{\mu}{\not{k}}P_{R}\}u(p_{b})
\varepsilon_{\mu}(k),
\end{eqnarray}
where the $C^{l}$ are coefficients that depend on $\hat{s},
\hat{t}$, and the masses, and the $f_{i}^{l}$ are form factors;
both the coefficients $C^{l}$ and the form factors $f_{i}^{l}$ are
given explicitly in Appendix A. The corresponding amplitude
squared is
\begin{equation}
\overline{\sum}|M_{ren}|^{2}=\overline{\sum}|M_{0}^{(s)}
+M_{0}^{(t)}|^{2} +2Re\overline{\sum}[(\sum_{l}\delta M^{l})
(M_{0}^{(s)} +M_{0}^{(t)})^{\dag}],
\end{equation}
where
\begin{eqnarray}
\overline{\sum}|M_{0}^{(s)} +M_{0}^{(t)}|^{2}&=&
\frac{g^{2}g_{s}^{2}}{2N_{C}m_{W}^{2}}
\{\frac{1}{(\hat{s}-m_{b}^{2})^{2}}[(m_{t}^{2}\cot^{2}\beta
+m_{b}^{2}\tan^{2}\beta)(p_{b}\cdot kp_{t}\cdot k \nonumber
\\ &-& m_{b}^{2}p_{t}\cdot k +2p_{b}\cdot kp_{b}\cdot
p_{t}-m_{b}^{2}p_{b}\cdot p_{t}) +2m_{b}^{2}m_{t}^{2}(p_{b}\cdot k
- m_{b}^{2})] \nonumber \\ &+&
\frac{1}{(\hat{t}-m_{t}^{2})^{2}}[(m_{t}^{2}\cot^{2}\beta
+m_{b}^{2}\tan^{2}\beta)(p_{b}\cdot kp_{t}\cdot k
+m_{t}^{2}p_{b}\cdot k \nonumber \\ &-& m_{t}^{2}p_{b}\cdot p_{t})
+2m_{b}^{2}m_{t}^{2}(p_{t}\cdot k -m_{t}^{2})]
+\frac{1}{(\hat{s}-m_{b}^{2})(\hat{t}-m_{t}^{2})} \nonumber \\
&\times& [(m_{t}^{2}\cot^{2}\beta
+m_{b}^{2}\tan^{2}\beta)(2p_{b}\cdot kp_{t}\cdot k +2p_{b}\cdot
kp_{b}\cdot p_{t} -2(p_{b}\cdot p_{t})^{2} \nonumber \\ &-&
m_{b}^{2}p_{t}\cdot k +m_{t}^{2}p_{b}\cdot k)
+2m_{b}^{2}m_{t}^{2}(p_{t}\cdot k -p_{b}\cdot k -2p_{b}\cdot
p_{t})]\},
\end{eqnarray}
\begin{eqnarray}
\overline{\sum}\delta M^{l}(M_{0}^{(s)})^{\dag} &=&
-\frac{g^{4}g_{s}^2}{64N_{C}\times 16\pi^{2}
m_{W}^{2}(\hat{s}-m_{b}^{2})}
C^{l}\sum_{i=1}^{12}h_{i}^{(s)}f_{i}^{l},
\end{eqnarray}
and
\begin{eqnarray}
\overline{\sum}\delta M^{l}(M_{0}^{(t)})^{\dag} &=&
-\frac{g^{4}g_{s}^2} {64N_{C}\times
16\pi^{2}m_{W}^{2}(\hat{t}-m_{t}^{2})} C^{l}\sum_{i=1}^{12}
h_{i}^{(t)}f_{i}^{l}.
\end{eqnarray}
Here the color factor $N_{C}=3$ and $h_{i}^{(s)}$ and
$h_{i}^{(t)}$ are scalar functions whose explicit expressions are
given in Appendix B.

The cross section for the process $gb\rightarrow tH^{-}$ is
\begin{equation}
\hat{\sigma} =\int_{\hat{t}_{min}}^{\hat{t}_{max}}\frac{1}{16\pi
\hat{s}^2} \overline{\Sigma}|M_{ren}|^{2}d\hat{t}
\end{equation}
with
\begin{eqnarray*}
\hat{t}_{min} &=& \frac{m_{t}^{2} +m_{H^{-}}^{2} -\hat{s}}{2}
-\frac{1}{2}\sqrt{(\hat{s} -(m_{t} +m_{H^{-}})^{2})(\hat{s}
-(m_{t} -m_{H^{-}})^{2})},
\end{eqnarray*}
and
\begin{eqnarray*}
\hat{t}_{max} &=& \frac{m_{t}^{2} +m_{H^{-}}^{2} -\hat{s}}{2}
+\frac{1}{2}\sqrt{(\hat{s} -(m_{t} +m_{H^{-}})^{2})(\hat{s}
-(m_{t} -m_{H^{-}})^{2})}.
\end{eqnarray*}
The total hadronic cross section for $pp\rightarrow gb\rightarrow
tH^{-}$ can be obtained by folding the subprocess cross section
$\hat{\sigma}$ with the parton luminosity:
\begin{equation}
\sigma(s) =\int_{(m_{t} +m_{H^{-}})/\sqrt{s}}^{1}dz \frac{dL}{dz}
\hat{\sigma}(gb\rightarrow tH^{-} \ \ {\rm at} \ \ \hat{s}
=z^{2}s).
\end{equation}
Here $\sqrt{s}$ and $\sqrt{\hat{s}}$ are the CM energies of the
$pp$ and $gb$ states , respectively, and $dL/dz$ is the parton
luminosity, defined as
\begin{equation}
\frac{dL}{dz} =2z\int_{z^{2}}^{1}
\frac{dx}{x}f_{b/P}(x,\mu)f_{g/P} (z^{2}/x,\mu),
\end{equation}
where $f_{b/P}(x,\mu)$ and $f_{g/P}(z^{2}/x,\mu)$ are the bottom
quark and gluon parton distribution functions. \vspace{.4cm}

\begin{center}{\Large 3. Numerical results and conclusion}\end{center}

In the following we present some numerical results for charged
Higgs boson production in association with a top quark at both the
Tevatron and the LHC. In our numerical calculations the SM
parameters were taken to be $m_W=80.41 GeV$, $m_Z=91.187 GeV$,
$m_t=176 GeV$, $\alpha_s(m_Z)=0.119$, and
$\alpha_{ew}(m_Z)={1\over 128.8}$[15]. And we used the running b
quark mass $\approx 3 GeV$ and the one-loop relations[16] from the
MSSM between the Higgs boson masses $m_{h,H,A,H^{\pm}}$ and the
parameters $\alpha$ and $ \beta$, and chose $m_{H^{\pm}}$ and
$\tan \beta$ as the two independent input parameters. And we used
the CTEQ5M[17] parton distributions throughout the calculations.
Other MSSM parameters were determined as follows:

(i) For the parameters $M_1,M_2$, and $\mu$ in the chargino and
neutralino matrix, we put $M_2=300GeV$ and then used the relation
$M_1=(5/3)(g'^2/g^2)M_2\simeq 0.5M_2$[2] to determine $M_1$. We
also put $\mu = -100 GeV$ except the numerical calculations as
shown in Fig.6(b), where $\mu$ is a variable.

(ii) For the parameters $m^2_{\tilde{Q},\tilde{U},\tilde{D}}$ and
$A_{t,b}$ in squark mass matrices

\begin{eqnarray}
M^2_{\tilde{q}} =\left(\begin{array}{cc} M_{LL}^2 & m_q M_{LR}\\
m_q M_{RL} & M_{RR}^2 \end{array} \right)
\end{eqnarray}
with
\begin{eqnarray}
M_{LL}^2 =m_{\tilde{Q}}^2 +m_q^2 +m_Z^2\cos 2\beta(I_q^{3L}
-e_q\sin^2\theta_W), \nonumber
\\ M_{RR}^2 =m_{\tilde{U},\tilde{D}}^2 +m_q^2 +m_Z^2
\cos 2\beta e_q\sin^2\theta_W, \nonumber
\\ M_{LR} =M_{RL} =\left(\begin{array}{ll} A_t -\mu\cot\beta &
(\tilde{q} =\tilde{t}) \\ A_b -\mu\tan\beta & (\tilde{q}
=\tilde{b}) \end{array} \right),
\end{eqnarray}
to simplify the calculation we assumed $m^2_{\tilde{Q}}
=m^2_{\tilde{U}} =m^2_{\tilde{D}}$ and $A_t=A_b$, and we put
$m_{\tilde{Q}}=500GeV$ and $A_t=200GeV$. But in the numerical
calculations of Fig.6(a) $A_t=A_b$ are the variables.

Some typical numerical calculations of the tree-level total cross
sections and the $O(\alpha_{ew}m^2_{t(b)}/m^2_W)$ SUSY electroweak
corrections as the functions of the charged Higgs boson mass,
$A_t=A_b$ and $\mu$, respectively, for three representative values
of $\tan\beta$ are given in Figs.3-6.

Figures 3(a) and 4(a) show that the tree-level total cross
sections as a function of the charged Higgs boson mass for three
representative values of $\tan\beta$. For $m_{H^{\pm}}=200GeV$ the
total cross sections at the Tevatron are at most only $0.7$ fb and
$0.1$ fb for $\tan\beta=2,30$ and $10$, respectively, and for
$m_{H^{\pm}}=300GeV$ the total cross sections are smaller than
$0.15$ fb for all three values of $\tan\beta$. However, at the LHC
the total cross sections are much larger: the order of thousands
of fb for $m_{H^{\pm}}$ in the range $100$ to $240 GeV$ and
$\tan\beta=2$ and $30$; and they are hundreds of fb for the
intermediate value $\tan\beta=10$. When the charged Higgs boson
mass becomes heavy($<500$ GeV), the total cross sections still are
larger than $100$ fb and $10$ fb for $\tan\beta=2,30$ and $10$,
respectively. For low $\tan\beta$ the top quark contribution is
enhanced while for high $\tan\beta$ the bottom quark contribution
becomes large. These results are smaller than ones given in
ref.[8,9] because we used the running b quark mass $\approx 3 GeV$
in the numerical calculations. We have confirmed that if we chose
$m_b=4.5 GeV$, our results will agree with ref.[8,9].

In Figs. 3(b) and 4(b) we show the corrections to the total cross
sections relative to the tree-level values as a function of
$m_{H^{\pm}}$ for $\tan\beta=2,10,$ and $30$. For $\tan\beta=2$
the corrections decrease the total cross sections significantly,
which exceed $-13\%$ for $m_{H^{\pm}}$ below $300GeV$ at the both
Tevatron and the LHC. But the corrections decrease as
$m_{H^{\pm}}$ increase. For example, as shown in Fig.4(b), the
corrections range between $-13\%\sim 0\%$ when $m_{H^{\pm}}$
increase from $300 GeV$ to $1 TeV$ at the LHC. For high
$\tan\beta(=10,30)$ these corrections become smaller, which can
decrease or increase the total cross sections depending on
$\tan\beta$, and the magnitude of the corrections are at most a
few percent for a wide range of the charged Higgs boson mass at
both the Tevatron and the LHC.

In Fig.5 we present the Yukawa correction from the Higgs sector
and the genuine SUSY electroweak correction from the couplings
involving the genuine SUSY particles(the chargino, neutralino and
squark) for $\tan\beta=30$ at the LHC, respectively. One can see
that the Yukawa correction and the genuine SUSY electroweak
correction have opposite signs, and thus cancel to some extent.
The former decrease the total cross sections, which can range
between $-8\%\sim -4\%$ for $m_{H^{\pm}}$ below $300GeV$, but the
latter increase the total cross sections, which range between
$10\%\sim 7\%$ for $m_{H^{\pm}}$ in the same range. In such a case
the combined effects just are about $2\%\sim 3\%$.

Figs.6(a) and 6(b) give the corrections as the functions of
$A_t=A_b$ and $\mu$ for $m_{H^{\pm}}=300 GeV$ at the LHC,
respectively, assuming $\tan\beta=2,10$ and $30$. From Figs.6(a)
and 6(b) one sees that the corrections increase or decrease slowly
with increasing $A_t=A_b$ and the magnitude of $\mu$,
respectively, for $\tan\beta=30,10$, and the corrections are not
very sensitive to both $A_t=A_b$ and $\mu$ for $\tan\beta=2$,
where the corrections are always about $-12\%$ and $-13\%$,
respectively. In general for large values of $A_t$ and small
values of $\tan\beta$ or large values of $\mu$ and $\tan\beta$,
one can get much larger corrections since the charged Higgs
boson-stop-sbottom couplings become stronger. For $\tan\beta=30$,
comparing Fig.4(b) with Fig.6(b), we can see that the corrections
indeed become larger as the values of $\mu$ increase. But for
$\tan\beta=2$ from Fig.4(a) and Fig.6(a) we found that the
corrections almost have no change when $A_t=A_b$ become larger.
Obviously the effects from the stronger couplings have been
suppressed by the decoupling effects because with an increase of
$A_t=A_b$ all the squark masses are still heavy, which almost is
same as discussed in Ref.[18].

In conclusion, we have calculated the
$O(\alpha_{ew}m_{t(b)}^{2}/m_{W}^{2})$ and $O(\alpha_{ew}
m_{t(b)}^4/m_W^4)$ SUSY electroweak corrections to the cross
section for the charged Higgs boson production in association with
a top quark at the Tevatron and the LHC. These corrections
decrease or increase the cross section depending on $\tan\beta$
but are not very sensitive to the mass of the charged Higgs boson
for high $\tan\beta$. At low $\tan\beta(=2)$ the corrections
decrease the total cross sections significantly, which exceed
$-12\%$ for $m_{H^{\pm}}$ below $300GeV$ at both the Tevatron and
the LHC, but for $m_{H^{\pm}}>300GeV$ the corrections can become
very small at the LHC. For high $\tan\beta(=10,30)$ these
corrections can decrease or increase the total cross sections, and
the magnitude of the corrections are at most a few percent 
at both the Tevatron and the LHC.

\vspace{.5cm}

This work was supported in part by the National Natural Science
Foundation of China, the Doctoral Program Foundation of Higher
Education of China, the Post Doctoral Foundation of China, a grant
from the State Commission of Science and Technology of China, and
the U.S.Department of Energy, Division of High Energy Physics,
under Grant No.DE-FG02-91-ER4086. S.H. Zhu also gratefully
acknowledges the support of the K.C. Wong Education Foundation of
Hong Kong. \eject

\begin{center}{\large Appendix A} \end{center}
\vspace{.7cm} The coefficients $C^l$ and form factors $f^l_i$ are
the following:
\begin{eqnarray*}
C^{V_{1}(s)} &=& \frac{m_{b}^{2}} {m_{W}^{2}(\hat{s}-m_{b}^{2})},
\ \ \ \ C^{V_{1}(t)}= \frac{m_{t}^{2}} {m_{W}^{2}(\hat{t}
-m_{t}^{2})},\ \ \ \ C^{s(s)} =\frac{m_{b}^{2}}{m_{W}^{2}(\hat{s}
-m_{b}^{2})^{2}}, \\ C^{s(t)} &=& \frac{m_{t}^{2}}{m_{W}^{2}
(\hat{t}-m_{t}^{2})^{2}}, \ \ \ \ C^{V_{2}(s)} = \frac{m_bm_t}
{m_W^2(\hat{s}-m_{b}^{2})},\hspace{1.4cm} C^{V_{2}(t)}
=\frac{m_bm_t} {m_W^2(\hat{t}-m_{t}^{2})}, \\ C^{b(s)} &=&
C^{b(t)} = \frac{m_tm_b}{m_W^2},
\end{eqnarray*}
\begin{eqnarray*}
   f_{1}^{V_{1}(s)} &=& \eta^{(1)}[m_{b}(g_{2}^{V_{1}(s)}
-g_{3}^{V_{1}(s)}) -2p_{b}\cdot k g_{6}^{V_{1}(s)}],
\\ f_{2}^{V_{1}(s)} &=& \eta^{(2)}[m_{b}(g_{3}^{V_{1}(s)}
-g_{2}^{V_{1}(s)}) -2p_{b}\cdot k g_{7}^{V_{1}(s)}],
\\ f_{3}^{V_{1}(s)} &=& \eta^{(2)}[2(g_{1}^{V_{1}(s)}
+g_{2}^{V_{1}(s)}) +m_{b}(g_{4}^{V_{1}(s)} +g_{5}^{V_{1}(s)})
+2p_{b}\cdot k g_{8}^{V_{1}(s)}],
\\ f_{4}^{V_{1}(s)} &=& \eta^{(1)}[2(g_{1}^{V_{1}(s)}
+g_{3}^{V_{1}(s)}) +m_{b}(g_{4}^{V_{1}(s)} +g_{5}^{V_{1}(s)})
+2p_{b}\cdot k g_{9}^{V_{1}(s)}],
\\ f_{7}^{V_{1}(s)} &=& \eta^{(2)} [-(g_{1}^{V_{1}(s)}
+g_{2}^{V_{1}(s)}) +m_{b}(g_{6}^{V_{1}(s)} +g_{7}^{V_{1}(s)})],
\\ f_{8}^{V_{1}(s)} &=& \eta^{(1)} [-(g_{1}^{V_{1}(s)}
+g_{3}^{V_{1}(s)}) +m_{b}(g_{6}^{V_{1}(s)} +g_{7}^{V_{1}(s)})],
\\ f_{9}^{V_{1}(s)} &=& \eta^{(1)} [g_{4}^{V_{1}(s)}
+2g_{6}^{V_{1}(s)} +m_{b}(g_{8}^{V_{1}(s)} -g_{9}^{V_{1}(s)})],
\\ f_{10}^{V_{1}(s)} &=& \eta^{(2)} [g_{5}^{V_{1}(s)}
+2g_{7}^{V_{1}(s)} +m_{b}(g_{9}^{V_{1}(s)} -g_{8}^{V_{1}(s)})],
\\ f_{1}^{V_{2}(s)} &=& 2p_{b}\cdot k g_{3}^{V_{2}(s)},
\hspace{3.4cm} f_{2}^{V_{2}(s)} = 2p_{b}\cdot k g_{4}^{V_{2}(s)},
\\ f_{3}^{V_{2}(s)} &=& 2g_{1}^{V_{2}(s)}
+2m_{t}\cot\beta(\delta\Lambda_{L}^{(1)} +\delta\Lambda_{L}^{(2)}
+\delta\Lambda_{L}^{(3)}) -2m_{t}g_{3}^{V_{2}(s)}
+2m_{b}g_{4}^{V_{2}(s)},
\\ f_{4}^{V_{2}(s)} &=& 2g_{2}^{V_{2}(s)}
+2m_{b}\tan\beta(\delta\Lambda_{R}^{(1)} +\delta\Lambda_{R}^{(2)}
+\delta\Lambda_{R}^{(3)}) +2m_{b}g_{3}^{V_{2}(s)}
-2m_{t}g_{4}^{V_{2}(s)},
\\ f_{7}^{V_{2}(s)} &=& -\frac{1}{2}f_{3}^{V_{2}(s)},
\hspace{4.0cm} f_{8}^{V_{2}(s)} = -\frac{1}{2}f_{4}^{V_{2}(s)},
\\ f_{1}^{V_{2}(t)} &=& 2p_{t}\cdot k g_{3}^{V_{2}(t)},
\hspace{3.6cm} f_{2}^{V_{2}(t)} = 2p_{t}\cdot k g_{4}^{V_{2}(t)},
\\ f_{5}^{V_{2}(t)} &=& 2g_{1}^{V_{2}(t)}
+2m_{t}\cot\beta(\delta\Lambda_{L}^{(1)} +\delta\Lambda_{L}^{(2)}
+\delta\Lambda_{L}^{(3)}) -2m_{t}g_{3}^{V_{2}(t)}
+2m_{b}g_{4}^{V_{2}(t)},
\\ f_{6}^{V_{2}(t)} &=& 2g_{2}^{V_{2}(t)} +2m_{b}\tan\beta
(\delta\Lambda_{R}^{(1)} +\delta\Lambda_{R}^{(2)}
+\delta\Lambda_{R}^{(3)}) +2m_{b}g_{3}^{V_{2}(t)}
-2m_{t}g_{4}^{V_{2}(t)},
\\ f_{7}^{V_{2}(t)} &=& -\frac{1}{2}f_{5}^{V_{2}(t)},
\hspace{4.0cm} f_{8}^{V_{2}(t)} = -\frac{1}{2}f_{6}^{V_{2}(t)},
\\ f_{1}^{s(s)} &=& 2\eta^{(1)}p_{b}\cdot k[g_{1}^{s(s)}
+m_{b}(g_{2}^{s(s)} +g_{3}^{s(s)})],
\\ f_{2}^{s(s)} &=& 2\eta^{(2)}p_{b}\cdot k[g_{5}^{s(s)}
+m_{b}(g_{2}^{s(s)} +g_{4}^{s(s)})],
\\ f_{3}^{s(s)} &=& 2\eta^{(2)}[m_{b}(g_{1}^{s(s)} +g_5^{s(s)})
+2(m_{b}^{2} +p_{b}\cdot k) g_{2}^{s(s)} +(m_{b}^{2} +2p_{b}\cdot
k) g_{3}^{s(s)} +m_{b}^{2}g_{4}^{s(s)}],
\\ f_{4}^{s(s)} &=& 2\eta^{(1)}[m_{b}(g_{1}^{s(s)} +g_5^{s(s)})
+2(m_{b}^{2} +p_{b}\cdot k) g_{2}^{s(s)} +m_{b}^{2}g_{3}^{s(s)}
+(m_{b}^{2} +2p_{b}\cdot k) g_{4}^{s(s)}],
\\ f_{7}^{s(s)} &=& -\frac{1}{2}f_{3}^{s(s)}, \hspace{4.0cm}
f_{8}^{s(s)} = -\frac{1}{2}f_{4}^{s(s)},
\\ f_{1}^{b(s)} &=& \sum_{i=H^{0},h^{0},G^{0},A^{0}}
\eta_{i}^{(3)}\{\eta^{(2)}[2m_{b}(-3D_{312} +(1-\zeta_{i})D_{27})
+m_{b}^{3}(D_{0} +D_{12} -D_{22}
  \\ & & -D_{32}) -m_{t}^{2}m_{b}(D_{23} +2D_{39}) -2m_{b}p_{b}
\cdot k(2D_{36} +D_{24} +\zeta_{i}(D_{0} +D_{12}))
  \\ & & +2m_{b}p_{t}\cdot k(D_{25} +D_{310}) +2m_{b}p_{b}\cdot
p_{t}(D_{26} +2D_{38})] +\eta^{(1)}[2m_{t}(-3D_{313} +(1
  \\ & & +\zeta_{i})D_{27}) -m_{t}^{3}(D_{33} +(1+\zeta_{i})
D_{23}) +m_{b}^{2}m_{t}(D_{13} -2D_{38} +(1 +\zeta_{i})(D_{0}
  \\ & & -D_{22})) +2m_{t}p_{b}\cdot k(D_{13} -D_{310}
-(1 +\zeta_{i})(D_{12} +D_{24})) +2m_{t}p_{t}\cdot k(2D_{37}
  \\ & & +(1 +\zeta_{i})D_{25}) +2m_{t}p_{b}\cdot p_{t}(2D_{39}
+(1+\zeta_{i})D_{26})]\}
  \\ & &(-k,-p_{b},p_{t},m_{b},m_{b},m_{i},m_{t})
  \\ & & -\frac{8\sqrt{2}m_W}{\sin{2\beta}}\sum_{i,j,k}N_{k4}N_{k3}
^{\ast}R_i(b)R_j(t)\sigma_{ij}D_{27}(-k,-p_b,p_t, m_{\tilde{b}_i},
m_{\tilde{b}_i},m_{\tilde\chi_k^0},m_{\tilde{t}_j}),
\\ f_{2}^{b(s)} &=& f_{1}^{b(s)}(\eta^{(1)}\leftrightarrow
\eta^{(2)}, L_l \leftrightarrow R_l, N_{kl} \leftrightarrow
N_{kl}^\ast),
\\ f_{3}^{b(s)} &=& \sum_{i=H^{0},h^{0},G^{0},A^{0}}
\eta_{i}^{(3)}\{\eta^{(1)}[-4D_{27} +2m_{b}^{2}(D_{22} -D_{0}
-(1-\zeta_{i})(D_{12} +D_{22}))
  \\ & & +2m_{t}^{2}(D_{23}-(1+\zeta_{i})D_{26}) +4p_{t}\cdot
k(D_{26} -D_{25})] +\eta^{(2)}2m_{t}m_{b}(1 +\zeta_{i})(D_{22}
  \\ & & -D_{12} -D_{26})\}(-k,-p_{b},p_{t},m_{b},m_{b},m_{i},m_{t})
  \\ & & -\frac{8\sqrt{2}m_W}{\sin{2\beta}}\sum_{i,j,k}\sigma_{ij}
[-m_tN_{k4}N_{k3}^{\ast}R_i(b)R_j(t)D_{26}
+m_bN_{k4}^{\ast}N_{k3}L_i(b)L_j(t)(D_{12}
  \\ & & +D_{22}) +m_{\tilde\chi_k^0}N_{k4}^{\ast}N_{k3}^{\ast}
R_i(b)L_j(t)D_{12}] (-k,-p_b,p_t,m_{\tilde{b}_i},
m_{\tilde{b}_i},m_{\tilde\chi_k^0},m_{\tilde{t}_j}),
\\ f_{4}^{b(s)} &=& f_{3}^{b(s)}(\eta^{(1)}\leftrightarrow
\eta^{(2)}, L_l \leftrightarrow R_l, N_{kl} \leftrightarrow
N_{kl}^\ast),
\\ f_{5}^{b(s)} &=& \sum_{i=H^{0},h^{0},G^{0},A^{0}}
\eta_{i}^{(3)}\{\eta^{(1)}[12D_{313} +2m_{b}^{2}(2D_{38} -D_{13}
+(1-\zeta_{i})(D_{13}
  \\ & & +D_{26})) +2m_{t}^{2}(D_{33}+(1+\zeta_{i})D_{23})
+4p_{b}\cdot k(D_{25} +D_{310}) -4p_{t}\cdot k(D_{23}
  \\ & & +2D_{37})-4p_{t}\cdot p_{b}(D_{23} +2D_{39})]
+\eta^{(2)}2m_{t}m_{b}(1 +\zeta_{i})(D_{13} +D_{23}
  \\ & & -D_{26})\}(-k,-p_{b},p_{t},m_{b},m_{b},m_{i},m_{t})
  \\ & & +\frac{8\sqrt{2}m_W}{\sin{2\beta}}\sum_{i,j,k}\sigma_{ij}
[-m_tN_{k4}N_{k3}^{\ast}R_i(b)R_j(t)D_{23}
+m_bN_{k4}^{\ast}N_{k3}L_i(b)L_j(t)(D_{13}
  \\ & & +D_{26})
+m_{\tilde\chi_k^0}N_{k4}^{\ast}N_{k3}^{\ast}R_i(b)L_j(t)D_{13}]
(-k,-p_b,p_t,m_{\tilde{b}_i},m_{\tilde{b}_i},m_{\tilde\chi_k^0},
m_{\tilde{t}_j}) ,
\\ f_{6}^{b(s)} &=& f_{5}^{b(s)}(\eta^{(1)}\leftrightarrow
\eta^{(2)}, L_l \leftrightarrow R_l, N_{kl} \leftrightarrow
N_{kl}^\ast),
\\ f_{7}^{b(s)} &=& \sum_{i=H^{0},h^{0},G^{0},A^{0}}
\eta_{i}^{(3)}\{\eta^{(1)}[6(D_{27} -D_{311}) +m_{b}^{2}(D_{11}
-2D_{12} -2D_{22}
  \\ & & -2D_{36} +(1+\zeta_{i})(D_{0} +D_{12}))-m_{t}^{2}(2D_{23}
+2D_{37} +(1+\zeta_{i})D_{13}) -2p_{b}\cdot k(D_{12}
  \\ & & +2D_{24} +2D_{34}) +2p_{t}\cdot k(D_{13} +2D_{25}
+2D_{35}) +2p_{t}\cdot p_{b}(D_{13} +2D_{26}
  \\ & & +D_{310})] +\eta^{(2)}m_{t}m_{b}(1 +\zeta_{i})(D_{12}
-D_{13} -D_{0})\}(-k,-p_{b},p_{t},m_{b},m_{b},m_{i},m_{t}),
\\ f_{8}^{b(s)} &=& f_{7}^{b(s)}(\eta^{(1)}\leftrightarrow
\eta^{(2)}),
\\ f_{9}^{b(s)} &=& \sum_{i=H^{0},h^{0},G^{0},A^{0}}
\eta_{i}^{(3)}\{\eta^{(1)}2m_{t}[-D_{13} -D_{26}
+(1+\zeta_{i})(D_{12} +D_{24})]
  \\ & & +\eta^{(2)}2m_{b}[-D_{22} +D_{24} +\zeta_{i}(D_{0}
+2D_{12} +D_{24})]\} (-k,-p_{b},p_{t},m_{b},m_{b},m_{i},m_{t})
  \\ & & -\frac{8\sqrt{2}m_W}{\sin{2\beta}}\sum_{i,j,k}\sigma_{ij}
N_{k4}N_{k3}^{\ast}R_i(b)R_j(t)(D_{12}
  \\ & & +D_{24})(-k,-p_b,p_t,m_{\tilde{b}_i},m_{\tilde{b}_i},
m_{\tilde\chi_k^0}, m_{\tilde{t}_j}),
\\ f_{10}^{b(s)} &=& f_{9}^{b(s)}(\eta^{(1)}\leftrightarrow
\eta^{(2)}, L_l \leftrightarrow R_l, N_{kl} \leftrightarrow
N_{kl}^\ast),
\\ f_{11}^{b(s)} &=& \sum_{i=H^{0},h^{0},G^{0},A^{0}}
\eta_{i}^{(3)}\{\eta^{(1)}2m_{t}[D_{23} -(1+\zeta_{i})D_{25}]
-\eta^{(2)}2m_{b}[-D_{26} +D_{25}
  \\ & & +\zeta_{i}(D_{13}
+D_{25})]\}(-k,-p_{b},p_{t},m_{b},m_{b},m_{i},m_{t})
  \\ & & +\frac{8\sqrt{2}m_W}{\sin{2\beta}}\sum_{i,j,k}\sigma_{ij}
N_{k4}N_{k3}^{\ast}R_i(b)R_j(t)(D_{13}
  \\ & & +D_{25})(-k,-p_b,p_t,m_{\tilde{b}_i},m_{\tilde{b}_i},
m_{\tilde\chi_k^0}, m_{\tilde{t}_j}),
\\ f_{12}^{b(s)} &=& f_{11}^{b(s)}(\eta^{(1)}\leftrightarrow
\eta^{(2)}, L_l \leftrightarrow R_l, N_{kl} \leftrightarrow
N_{kl}^\ast),
\end{eqnarray*}
where $D_{0},D_{ij},D_{ijk}$ are the four-point Feynman integrals
[19]. The explicit forms of $\delta M^{V_1(t)}, \delta M^{s(t)},
\delta M^{b(t)}$ can be respectively obtained from $\delta
M^{V_1(s)}, \delta M^{s(s)}, \delta M^{b(s)}$ by the
transformation $U$ defined as $$p_{b}\rightarrow p_{t},\ \ \ \
\hat{s}\rightarrow \hat{t},\ \ \ \ k\rightarrow -k, \ \ \ \
\xi_{i}^{(1)}\rightarrow \xi_{i}^{(2)},\ \ \ \
\xi_{i}^{(3)}\rightarrow \xi_{i}^{(4)},\ \ \ \ \eta_i^{(1)}
\rightarrow \eta_i^{(2)},$$ $$m_{t}\leftrightarrow m_{b},\ \ \
\eta^{(1)}\leftrightarrow\eta^{(2)},\ \ \ \lambda_b
\leftrightarrow \lambda_t,\ \ \  m_{\tilde{t}_i} \leftrightarrow
m_{\tilde{b}_i},\ \ \  U_{i2} \leftrightarrow V_{i2}^\ast,\ \ \
N_{i3} \leftrightarrow N_{i4}^\ast,$$ $$L_i(b) \leftrightarrow
L_i(t), \ \ \ R_i(b) \leftrightarrow R_i(t), \ \ \ p_b^\mu
P_{L(R)} \leftrightarrow p_t^\mu P_{R(L)}, \ \ \ \gamma^\mu
\not{k}P_L \leftrightarrow \gamma^\mu \not{k}P_R.$$

All other form factors $f_i^l$ not listed above vanish. In the
above expressions we have used the following definitions:
\begin{eqnarray*}
\eta^{(1)} =m_{b}\tan\beta,\hspace{1.0cm} \eta^{(2)}
=m_{t}\cot\beta, & & \lambda_b =\frac{m_b}{\sqrt{2}m_W\cos\beta},
\hspace{1.0cm} \lambda_t =\frac{m_t}{\sqrt{2}m_W\sin\beta}
\\ L_1(q) =\cos\theta_q,\hspace{1.0cm} L_2(q) =-\sin\theta_q, & &
R_1(q) =\sin\theta_q,\hspace{1.0cm} R_2(q) =\cos\theta_q,
\\ \eta_{H^{0}}^{(1)} =\frac{\cos^{2}\alpha}{\cos^{2}\beta},
\hspace{1.5cm} \eta_{h^{0}}^{(1)} =\frac{\sin^{2}\alpha}
{\cos^{2}\beta}, & &\hspace{0.1cm} \eta_{A^{0}}^{(1)}
=\tan^{2}\beta,\hspace{1.5cm} \eta_{G^{0}}^{(1)} =1,
\\ \eta_{H^{0}}^{(2)} =\frac{\sin^{2}\alpha}{\sin^{2}\beta},
\hspace{1.5cm} \eta_{h^{0}}^{(2)} =\frac{\cos^{2}\alpha}
{\sin^{2}\beta},& & \hspace{0.1cm} \eta_{A^{0}}^{(2)}
=\cot^{2}\beta, \hspace{1.6cm} \eta_{G^{0}}^{(2)}=1,
\\ \eta_{H^{0}}^{(3)} =-\eta_{h^{0}}^{(3)} =\frac{\sin\alpha
\cos\alpha}{\sin\beta \cos\beta},\hspace{1.7cm} & &
\eta_{G^{0}}^{(3)} =-\eta_{A^{0}}^{(3)} =1,
\\ \xi_{H^{-}}^{(1)} =\frac{m_{t}^{2}}{m_{b}^{2}} \cot^{2}\beta,
\hspace{1.2cm}\xi_{G^{-}}^{(1)} =\frac{m_{t}^{2}}{m_{b}^{2}}, & &
\hspace{0.2cm}\xi_{H^{-}}^{(2)}
=\frac{m_{b}^{2}}{m_{t}^{2}}\tan^{2}\beta, \hspace{1.0cm}
\xi_{G^{-}}^{(2)} =\frac{m_{b}^{2}}{m_{t}^{2}},
\\ \xi_{H^{-}}^{(3)} =\tan^{2}\beta, \hspace{1.7cm}
\xi_{G^{-}}^{(3)} =1,\hspace{0.6cm} & &\hspace{0.2cm}
\xi_{H^{-}}^{(4)} =\cot^{2}\beta, \hspace{1.5cm} \xi_{G^{-}}^{(4)}
=1,
\end{eqnarray*}
$$\zeta_{H^{0}} = \zeta_{h^{0}} =\zeta_{H^{-}} =-\zeta_{A^{0}}
=-\zeta_{G^{0}} =-\zeta_{G^{-}} =1,$$
\begin{eqnarray*}
\sigma_{ij} &=& \frac{m_W}{\sqrt{2}}(\sin{2\beta}
-\frac{m_b^2\tan\beta +m_t^2\cot\beta}{m_W^2})L_i(b)L_j(t)
  \\ & & +\frac{m_tm_b}{\sqrt{2}m_W}(\tan\beta
+\cot\beta)R_i(b)R_j(t) -\frac{m_b}{\sqrt{2}m_W}(\mu
-A_b\tan\beta)R_i(b)L_j(t)
  \\ & & -\frac{m_t}{\sqrt{2}m_W}(\mu -A_t\cot\beta)L_i(b)R_j(t),
\\ g_{1}^{V_{1}(s)} &=& \sum_{i=H^{0},h^{0},G^{0},A^{0}}
\eta_{i}^{(1)}\{[\frac{1}{2} -2\overline{C}_{24} +m_{b}^{2}
(-2C_{11} +C_{12} -C_{21}+C_{23}) -\hat{s}(C_{12}
  \\ & & +C_{23})](-p_{b},-k,m_{i},m_{b},m_{b}) +[-F_{0} +F_{1}
+2m_{b}^{2}G_{1}
  \\ & & -(1+\zeta_{i})2m_{b}^{2}G_{0}](m_{b}^{2},m_{i},m_{b})\},
\\ g_{2}^{V_{1}(s)} &=& \sum_{i=H^{-},G^{-}} 2\{\xi_{i}^{(1)}
[\frac{1}{2} -2\overline{C}_{24} +m_{t}^{2}C_{0} +m_{b}^{2}(-C_{0}
-2C_{11} +C_{12} -C_{21} +C_{23})
  \\ & & -\hat{s}(C_{12} +C_{23})](-p_{b},-k,m_{i},m_{t},m_{t})
+[\xi_{i}^{(1)}(-F_{0} +F_{1}) -2m_{t}^{2}\zeta_{i}G_{0}
  \\ & & +m_{b}^{2}(\xi_{i}^{(1)}
+\xi_{i}^{(3)})(G_{1} -\zeta_{i}G_{0})](m_{b}^{2},m_{i},m_{t})\}
  \\ & & +\frac{4m_W^2}{m_b^2}\sum_{i,j}\{\lambda_b^2[R_j^2(b)
|N_{i3}|^2(-F_0 +F_1) +m_b^2|N_{i3}|^2(-G_0 +G_1)
-2m_bm_{\tilde\chi_i^0}
  \\ & & \times L_j(b)R_j(b)N_{i3}^{\ast2}G_0]
(m_b^2,m_{\tilde{b}_j},m_{\tilde\chi_i^0})
+[-2m_bm_{\tilde\chi_i^+}\lambda_b\lambda_t
L_j(t)R_j(t)V_{i2}^{\ast2}U_{i2}^{\ast2}G_0
  \\ & & +\lambda_t^2R_j^2(t)|V_{i2}|^2(-F_0 +F_1)
+m_b^2(\lambda_t^2R_j^2(t)|V_{i2}|^2
+\lambda_b^2L_j^2(t)|U_{i2}|^2)(-G_0
  \\ & & +G_1)](m_b^2,m_{\tilde{t}_j},m_{\tilde\chi_i^+})
-2\lambda_b^2R_j^2(b)|N_{i3}|^2\bar{C}_{24}
(-p_b,-k,m_{\tilde\chi_i^0},m_{\tilde{b}_j},m_{\tilde{b}_j})
  \\ & & -2\lambda_t^2R_j^2(t)|V_{i2}|^2\bar{C}_{24}
(-p_b,-k,m_{\tilde\chi_i^+},m_{\tilde{t}_j},m_{\tilde{t}_j})\} ,
\\ g_{3}^{V_{1}(s)} &=& g_{2}^{V_{1}(s)}(\xi_{i}^{(1)} \leftrightarrow
\xi_{i}^{(3)}, V_{i2} \leftrightarrow U_{i2}^\ast, N_{i3}
\leftrightarrow N_{i3}^\ast, L_j(b) \leftrightarrow R_j(b),
\lambda_bL_j(t) \leftrightarrow \lambda_tR_j(t)),
\\ g_{4}^{V_{1}(s)} &=& \sum_{i=H^{0},h^{0},G^{0},A^{0}}
\eta_{i}^{(1)}2m_{b} [C_{0} +2C_{11} +C_{21} +\zeta_{i}(C_{0}
+C_{11})] (-p_{b},-k,m_{i},m_{b},m_{b})
  \\ & & +\sum_{i=H^{-},G^{-}} 4m_{b}[\xi_{i}^{(3)}(C_0 +2C_{11}
+C_{21}) +\frac{m_{t}^{2}} {m_{b}^{2}}\zeta_{i}(C_{0} +C_{11})]
(-p_{b},-k,m_{i},m_{t},m_{t})
  \\ & & +\frac{8m_W^2}{m_b^2}\sum_{i,j}\{\lambda_b^2
[m_{\tilde\chi_i^0}L_j(b)R_j(b)N_{i3}^{\ast2} (C_0 +C_{11})
-m_bL_j^2(b)|N_{i3}|^2 (C_{11}
  \\ & & +C_{21})](-p_b,-k,m_{\tilde\chi_i^0},m_{\tilde{b}_j},
m_{\tilde{b}_j})
  \\ & & +[m_{\tilde\chi_i^+}\lambda_b \lambda_tL_j(t)R_j(t)
V_{i2}^{\ast}U_{i2}^{\ast}(C_0 +C_{11}) -m_b\lambda_b^2L_j^2(t)
|U_{i2}|^2 (C_{11}
  \\ & & +C_{21})](-p_b,-k,m_{\tilde\chi_i^+},m_{\tilde{t}_j},
m_{\tilde{t}_j})\} ,
\\ g_{5}^{V_{1}(s)} &=& g_{4}^{V_{1}(s)}(\xi_{i}^{(1)} \leftrightarrow
\xi_{i}^{(3)}, V_{i2} \leftrightarrow U_{i2}^\ast, N_{i3}
\leftrightarrow N_{i3}^\ast, L_j(b) \leftrightarrow R_j(b),
\lambda_bL_j(t) \leftrightarrow \lambda_tR_j(t)),
\\ g_{6}^{V_{1}(s)}
&=& -\sum_{i=H^{0},h^{0},G^{0},A^{0}} \eta_{i}^{(1)}m_{b} (C_{0}
+C_{11} +\zeta_{i}C_{0})(-p_{b},-k,m_{i},m_{b},m_{b})
  \\ & & -\sum_{i=H^{-},G^{-}} 2m_{b}[\xi_{i}^{(3)}(C_{0} +C_{11})
+\frac{m_{t}^{2}} {m_{b}^{2}}\zeta_{i}C_{0}]
(-p_{b},-k,m_{i},m_{t},m_{t}),
\\ g_{7}^{V_{1}(s)} &=& g_{6}^{V_{1}(s)}(\xi_{i}^{(1)} \leftrightarrow
\xi_{i}^{(3)}),
\\ g_{8}^{V_{1}(s)} &=& \sum_{i=H^{0},h^{0},G^{0},A^{0}}
2\eta_{i}^{(1)}(C_{12} +C_{23}) (-p_{b},-k,m_{i},m_{b},m_{b})
  \\ & & +\sum_{i=H^{-},G^{-}} 4\xi_{i}^{(1)}(C_{12} +C_{24})
(-p_{b},-k,m_{i},m_{t},m_{t})
  \\ & & -\frac{8m_W^2}{m_b^2}\sum_{i,j}\{\lambda_b^2R_j^2(b)
|N_{i3}|^2(C_{12} +C_{23})
(-p_b,-k,m_{\tilde\chi_i^0},m_{\tilde{b}_j},m_{\tilde{b}_j})
  \\ & & +\lambda_t^2R_j^2(t)|V_{i2}|^2(C_{12} +C_{23})
(-p_b,-k,m_{\tilde\chi_i^+},m_{\tilde{t}_j},m_{\tilde{t}_j})\},
\\ g_{9}^{V_{1}(s)} &=& g_{8}^{V_{1}(s)}(\xi_{i}^{(1)} \leftrightarrow
\xi_{i}^{(3)}, V_{i2} \leftrightarrow U_{i2}^\ast, N_{i3}
\leftrightarrow N_{i3}^\ast, L_j(b) \leftrightarrow R_j(b),
\lambda_bL_j(t) \leftrightarrow \lambda_tR_j(t)),
\\ g_{1}^{V_{2}(s)} &=& \sum_{i=H^{0},h^{0},G^{0},A^{0}}
\eta_{i}^{(3)}\{\eta^{(1)}[-\frac{1}{2} +4\overline{C}_{24}
+m_{t}^{2}(C_{0} +2C_{11} +\zeta_{i}(C_{0} +C_{11})
  \\ & & +C_{21} -C_{12} -C_{23}) +m_{H^{-}}^{2}(C_{22} -C_{23})
+\hat{s}(C_{12} +C_{23})] +\eta^{(2)}m_{b}m_{t}[\zeta_{i}C_{11}
  \\ & & +(1 +\zeta_i)C_{0}]\}(-p_{t},-p_{H^{-}},m_{i},
m_{t},m_{b}) +\frac{4\sqrt{2}m_W}{\sin{2\beta}}\sum_{i,j,k}
[m_tR_i(b)R_j(t) N_{k3}^{\ast}N_{k4}
  \\ & & \times (-C_{11} +C_{12})+m_{\tilde\chi_k^0}L_j(t)R_i(b)
N_{k3}^{\ast}N_{k4}^{\ast} C_0]\sigma_{ij}(-p_t,-p_{H^-},
m_{\tilde\chi_k^0},m_{\tilde{b}_i}, m_{\tilde{t}_j}),
\\ g_{2}^{V_{2}(s)} &=& g_{1}^{V_{2}(s)}(\eta^{(1)}\leftrightarrow
\eta^{(2)}, L_l \leftrightarrow R_l, N_{kl} \leftrightarrow
N_{kl}^\ast),
\\ g_{3}^{V_{2}(s)} &=& \sum_{i=H^{0},h^{0},G^{0},A^{0}}
\eta_{i}^{(3)}\{\eta^{(1)}m_{t}[C_{0} +C_{11} +\zeta_{i}(C_{0}
+C_{12})] +\eta^{(2)}\zeta_{i} m_{b}C_{12}\}
  \\ & & (-p_{t},-p_{H^{-}},m_{i},m_{t},m_{b})
  \\ & & -\frac{4\sqrt{2}m_W}{\sin{2\beta}}\sum_{i,j,k}R_i(b)R_j(t)
N_{k3}^{\ast}N_{k4}\sigma_{ij}C_{12} (-p_t,-p_{H^-},
m_{\tilde\chi_k^0}, m_{\tilde{b}_i},m_{\tilde{t}_j}),
\\ g_{4}^{V_{2}(s)} &=& g_{3}^{V_{2}(s)}(\eta^{(1)}\leftrightarrow
\eta^{(2)}, L_l \leftrightarrow R_l, N_{kl} \leftrightarrow
N_{kl}^\ast),
\\ g_{1}^{V_{2}(t)} &=& \sum_{i=H^{0},h^{0},G^{0},A^{0}}
\eta_{i}^{(3)}\{\eta^{(1)}[-\frac{1}{2} +4\overline{C}_{24}
+m_{b}^{2}(C_{0} +2C_{11} +\zeta_{i}(C_{0} +C_{11})
  \\ & & +C_{21} -C_{12} -C_{23}) +m_{H^{-}}^{2}(C_{22}
-C_{23}) +\hat{t}(C_{12} +C_{23})]
  \\ & & +\eta^{(2)}m_{b}m_{t}[C_{0} +\zeta_{i}(C_{0} +C_{11})]\}
(-p_{b},p_{H^{-}},m_{i},m_{b},m_{t})
  \\ & & +\frac{4\sqrt{2}m_W}{\sin{2\beta}}\sum_{i,j,k}
[m_bL_i(b)L_j(t)N_{k3}^{\ast}N_{k4}(-C_{11} +C_{12})
  \\ & & +m_{\tilde\chi_k^0}L_j(t)R_i(b)N_{k3}^{\ast}N_{k4}^{\ast}
C_0]\sigma_{ij}(-p_b,p_{H^-},m_{\tilde\chi_k^0},
m_{\tilde{b}_i},m_{\tilde{t}_j}),
\\ g_{2}^{V_{2}(t)} &=&
g_{1}^{V_{2}(t)}(\eta^{(1)}\leftrightarrow \eta^{(2)}, L_l
\leftrightarrow R_l, N_{kl} \leftrightarrow N_{kl}^\ast),
\\ g_{3}^{V_{2}(t)} &=& -\sum_{i=H^{0},h^{0},G^{0},A^{0}}
\eta_{i}^{(3)}\{\eta^{(1)}m_{b}[C_{0} +C_{11} +\zeta_{i}(C_{0}
+C_{12})] +\eta^{(2)}\zeta_{i} m_{t}C_{12}\}
  \\ & & (-p_{b},p_{H^{-}},m_{i},m_{b},m_{t})
  \\ & & +\frac{4\sqrt{2}m_W}{\sin{2\beta}}\sum_{i,j,k}R_i(b)R_j(t)
N_{k3}^{\ast}N_{k4}\sigma_{ij}C_{12} (-p_b,p_{H^-},
m_{\tilde\chi_k^0}, m_{\tilde{b}_i},m_{\tilde{t}_j}) ,
\\ g_{4}^{V_{2}(t)} &=& g_{3}^{V_{2}(t)}(\eta^{(1)}\leftrightarrow
\eta^{(2)}, L_l \leftrightarrow R_l, N_{kl} \leftrightarrow
N_{kl}^\ast),
\\ g_{1}^{s(s)} &=& \sum_{i=H^{0},h^{0},G^{0},A^{0}}m_{b}
\eta_{i}^{(1)}\{-\zeta_{i}F_{0}(p_{b}+k,m_{i},m_{b})
+[\zeta_{i}F_{0} -2m_{b}^{2}(1 +\zeta_{i})G_{0}
  \\ & & +2m_{b}^{2}G_{1}](m_{b}^{2},m_{i},m_{b})\}
+\sum_{i=H^{-},G^{-}}2m_{b}\{-\frac{m_{t}^{2}}{m_{b}^{2}}\zeta_{i}
F_{0}(p_{b}+k,m_{i},m_{t})
  \\ & & +[-2m_{t}^{2}\zeta_{i}G_{0} +m_{b}^{2} (\xi_{i}^{(1)}
+\xi_{i}^{(3)})(G_{1} -\zeta_{i}G_{0}) +\zeta_{i}
\frac{m_{t}^{2}}{m_{b}^{2}}F_{0}] (m_{b}^{2},m_{i},m_{t})\}
  \\ & & +\frac{4m_W^2}{m_b^2}\sum_{i,j}\{-m_{\tilde\chi_i^0}
\lambda_b^2 L_j(b)R_j(b)N_{i3}^{\ast2}F_0(p_b+k,m_{\tilde{b}_j},
m_{\tilde\chi_i^0})
  \\ & & -m_{\tilde\chi_i^+}\lambda_b\lambda_tL_j(b)R_j(b)
V_{i2}^{\ast} U_{i2}^{\ast}F_0(p_b+k,m_{\tilde{t}_j},
m_{\tilde\chi_i^+}) +[m_b^3\lambda_b^2|N_{i3}|^2(-G_0 +G_1)
  \\ & & -m_{\tilde\chi_i^0}\lambda_b^2
L_j(b)R_j(b)N_{i3}^{\ast2}(2m_b^2G_0 -F_0)](m_b^2,m_{\tilde{b}_j},
m_{\tilde\chi_i^0}) +[m_b^3(\lambda_b^2L_j^2(t)|U_{i2}|^2
  \\ & & +\lambda_t^2R_j^2(t)|V_{i2}|^2)(-G_0 +G_1)
-m_{\tilde\chi_i^+}\lambda_b\lambda_t
L_j(t)R_j(t)V_{i2}^{\ast}U_{i2}^{\ast}(2m_b^2G_0
  \\ & & -F_0)](m_b^2,m_{\tilde{t}_j}, m_{\tilde\chi_i^+})\},
\\ g_{2}^{s(s)} &=& \sum_{i=H^{0},h^{0},G^{0},A^{0}}
\eta_{i}^{(1)}(-F_{0} +F_{1})(p_{b}+k,m_{i},m_{b}),
\\ g_{3}^{s(s)} &=& \sum_{i=H^{0},h^{0},G^{0},A^{0}}\eta_{i}^{(1)}
[F_{0} -F_{1} -2m_{b}^{2}G_{1} +2(1 +\zeta_{i})m_{b}^{2}G_{0}]
(m_{b}^{2},m_{i},m_{b})
  \\ & & +\sum_{i=H^{-},G^{-}}2\{\xi_{i}^{(1)}(-F_{0} +F_{1})
(p_{b}+k,m_{i},m_{t}) -[\xi_{i}^{(1)}(-F_{0}+F_{1})
  \\ & &-2\zeta_{i}m_{t}^{2}G_{0} +m_{b}^{2}(\xi_{i}^{(1)}
+\xi_{i}^{(3)}) (G_{1} -\zeta_{i}G_{0})](m_{b}^{2},m_{i},m_{t})\}
  \\ & & -\frac{4m_W^2}{m_b^2}\sum_{i,j}\{\lambda_b^2[R_j^2(b)
|N_{i3}|^2(-F_0 +F_1) +|N_{i3}|^2m_b^2(-G_0 +G_1)
  \\ & & -2m_bm_{\tilde\chi_i^0}L_j(b)R_j(b)N_{i3}^{\ast2}G_0]
(m_b^2,m_{\tilde b_j},m_{\tilde\chi_i^0})
  \\ & & +[\lambda_t^2R_j^2(t)|V_{i2}|^2(-F_0 +F_1)
+m_b^2(\lambda_t^2R_j^2(t)|V_{i2}|^2
+\lambda_b^2L_j^2(t)|U_{i2}|^2)(G_1 -G_0)
  \\ & & -2m_bm_{\tilde\chi_i^+}L_j(t)R_j(t)\lambda_b
\lambda_tV_{i2}^{\ast} U_{i2}^{\ast} G_0] (m_b^2,m_{\tilde
t_j},m_{\tilde\chi_i^+})
  \\ & & -\lambda_b^2R_j^2(b)|N_{i3}|^2 (-F_0 +F_1)
(p_b+k,m_{\tilde{b}_j}, m_{\tilde\chi_i^0})
  \\ & & -\lambda_t^2R_j^2(t)|V_{i2}|^2 (-F_0
+F_1)(p_b+k,m_{\tilde{t}_j}, m_{\tilde\chi_i^+})\},
\\ g_{4}^{s(s)} &=& g_{3}^{s(s)}(\xi_{i}^{(1)} \leftrightarrow
\xi_{i}^{(3)}, V_{i2} \leftrightarrow U_{i2}^\ast, N_{i3}
\leftrightarrow N_{i3}^\ast, L_j(b) \leftrightarrow R_j(b),
\lambda_bL_j(t) \leftrightarrow \lambda_tR_j(t)),
\\ g_{5}^{s(s)} &=& g_{1}^{s(s)}(N_{i3}^{\ast} \rightarrow N_{i3},
V_{i2}^{\ast} \rightarrow V_{i2},U_{i2}^{\ast} \rightarrow
U_{i2}),
\\ \delta\Lambda_{L}^{(1)} &=& \frac{4N_{c}} {3m_{W}^{2}}
(1-\cot^{2}\theta_{W})[2m_{t}^{2}(\ln{\frac{m_{t}^{2}}{\mu^{2}}}
-1) +m_{b}^{2} +m_{t}^{2} -\frac{5}{6}m_{W}^{2} +m_{b}^{2}F_{0}
  \\ & & +(m_{b}^{2}-m_{t}^{2}-2m_{W}^{2})F_{1}]
(m_{W}^{2},m_{b},m_{t}) +\frac{4N_{c}}{3m_{W}^{2}} \cot^{2}
\theta_{W}\{-\frac{5}{6}[(g_{V}^{b})^{2} +(g_{A}^{b})^{2}
  \\ & & +(g_{V}^{t})^{2} +(g_{A}^{t})^{2}]m_{Z}^{2}
+[((g_{V}^{t})^{2} +(g_{A}^{t})^{2})(2m_{t}^{2}
\ln{\frac{m_{t}^{2}}{\mu^{2}}} +m_{t}^{2}F_{0} -2m_{Z}^{2}F_{1})
  \\ & & -((g_{V}^{t})^{2} -(g_{A}^{t})^{2})3m_{t}^{2}F_{0}]
(m_{Z}^{2},m_{t},m_{t}) +[((g_{V}^{b})^{2} +(g_{A}^{b})^{2})
(2m_{b}^{2}\ln{\frac{m_{b}^{2}} {\mu^{2}}}
  \\ & & +m_{b}^{2}F_{0} -2m_{Z}^{2}F_{1})-((g_{V}^{b})^{2}
-(g_{A}^{b})^{2}) 3m_{b}^{2}F_{0}] (m_{Z}^{2},m_{b},m_{b})\}
+\frac{4N_{c}}{m_{W}^{2}} [(\cot^{2}\beta
  \\ & & -1)m_{t}^{2}F_{0} +(m_{t}^{2}-m_{b}^{2}
-2m_{t}^{2}\cot^{2}\beta)F_{1} +(m_{t}^{2}\cot^{2}\beta +m_{b}^{2}
\tan^{2}\beta
  \\ & & +2m_{b}^{2})m_{t}^{2}G_{0} -(m_{t}^{2}\cot^{2}\beta
+m_{b}^{2} \tan^{2}\beta)m_{H^{-}}^{2}G_{1}] (m_{H^{-}}^{2},m_{t},
m_{b})
  \\ & & +\sum_{i=H^{0},h^{0},G^{0},A^{0}} \frac{1}{2m_{W}^{2}}
\{m_{b}^{2}\eta_{i}^{(1)}[F_{1}-F_{0} -2m_{b}^{2}(G_0
+\zeta_{i}G_{0} -G_{1})] (m_{b}^{2},m_{i},m_{b})
  \\ & & -m_{t}^{2} \eta_{i}^{(2)}[-(1+2\zeta_{i})F_{0} +F_{1}
+2m_{t}^{2} (1+\zeta_{i})G_{0} -2m_{t}^{2}G_{1}]
(m_{t}^{2},m_{i},m_{t})\}
  \\ & & +\sum_{i=H^{-},G^{-}}\frac{1}{m_{W}^{2}}
\{m_{b}^{2}[\xi_{i}^{(1)}(-F_{0} +F_{1}) -2m_{t}^{2}\zeta_{i}G_{0}
+m_{b}^{2}(\xi_{i}^{(1)} +\xi_{i}^{(3)})
  \\ & & \times(G_{1} -\zeta_{i}G_{0})](m_{b}^{2},m_{i},m_{t})
-m_{t}^{2}[-\frac{2m_{b}^{2}}{m_{t}^{2}}\zeta_{i}F_{0}
+\xi_{i}^{(2)}(-F_{0} +F_{1}) +2m_{b}^{2}\zeta_{i}G_{0}
  \\ & & -m_{t}^{2}(\xi_{i}^{(2)} +\xi_{i}^{(4)})(G_{1}
-\zeta_{i}G_{0})](m_{t}^{2},m_{i},m_{b})\}
-2N_C\sum_{i,j}\{2\sigma_{ij}\sigma_{ij}G_0
  \\ & & +\frac{1}{m_W^2}L_i(b)L_j(t)[L_i(b)L_j(t) (\frac{m_b^2}
{\cos^2\beta} +\frac{m_t^2}{\sin^2\beta})\cos{2\beta}
  \\ & & +R_i(b)R_j(t)m_tm_b(\tan^2\beta
-\cot^2\beta)]\}(m_{H^-}^2,m_{\tilde{t}_j}, m_{\tilde{b}_i}) ,
\\ \delta\Lambda_{L}^{(2)} &=& -2\sum_{i,j}\{\lambda^2_t[-\frac
{2m_{\tilde{\chi}^0_i}}{m_t} L_j(t)R_j(t)N_{i4}^{\ast2}(F_0
-m_t^2G_0) +|N_{i4}|^2(R_j^2(t)(-F_0 +F_1)
  \\ & & -m_t^2(-G_0 +G_1))] (m_t^2,m_{\tilde{t}_j},
m_{\tilde{\chi}_i^0}) +[-\frac {2m_{\tilde{\chi}^+_i}}{m_t}
\lambda_b\lambda_t L_j(b)R_j(b)U_{i2}^\ast V_{i2}^\ast(F_0
  \\ & & -m_t^2G_0) +\lambda_b^2R_j^2(b)|U_{i2}|^2(-F_0 +F_1)
-m_t^2(\lambda_t^2L_j^2(b)|V_{i2}|^2
+\lambda_b^2R_j^2(b)|U_{i2}|^2)
  \\ & & \times(-G_0 +G_1)](m_t^2,m_{\tilde{b}_j},
m_{\tilde{\chi}_i^+})\},
\\ \delta\Lambda_{L}^{(3)} &=& 2\sum_{i,j}\{\lambda^2_b
[|N{i3}|^2(R_j^2(b)(-F_0 +F_1) +m_b^2(-G_0 +G_1))
-2m_bm_{\tilde{\chi}_i^0}L_j(b)R_j(b)
  \\ & & \times N_{i3}^{\ast2}G_0] (m_b^2,m_{\tilde{b}_j},
m_{\tilde{\chi}_i^0}) +[-2m_{\tilde{\chi}^+_i} m_b
\lambda_b\lambda_t L_j(t)R_j(t)U_{i2}^\ast V_{i2}^\ast G_0
  \\ & & +\lambda_b^2L_j^2(b)|U_{i2}|^2(-F_0 +F_1) +m_b^2
(\lambda_t^2R_j^2(t)|V_{i2}|^2 +\lambda_b^2L_j^2(t)|U_{i2}|^2)
  \\ & & \times(-G_0 +G_1)]
(m_b^2,m_{\tilde{t}_j},m_{\tilde{\chi}_i^+})\},
\\ \delta\Lambda_{R}^{(1)} &=& \sum_{i=H^{0},h^{0},G^{0},A^{0}}
\frac{1}{2m_{W}^{2}} \{m_{t}^{2}\eta_{i}^{(2)}[-F_{0} +F_{1}
-2m_{t}^{2}(G_0+\zeta_{i}G_{0} -G_{1})] (m_{t}^{2},m_{i},m_{t})
  \\ & & -m_{b}^{2} \eta_{i}^{(1)}[-F_{0} +F_{1} -2\zeta_{i}F_{0}
+2m_{b}^{2}(1+\zeta_{i})G_{0} -2m_{b}^{2}G_{1}]
(m_{b}^{2},m_{i},m_{b})\}
  \\ & & +\sum_{i=H^{-},G^{-}}\frac{1}{m_{W}^{2}}\{m_{t}^{2}
[\xi_{i}^{(2)} (-F_{0} +F_{1}) -2m_{b}^{2}\zeta_{i}G_{0}
+m_{t}^{2}(\xi_{i}^{(2)} +\xi_{i}^{(4)})(G_{1}
  \\ & & -\zeta_{i}G_{0})](m_{t}^{2},m_{i},m_{b})
-m_{b}^{2}[-\frac{2m_{t}^{2}}{m_{b}^{2}}\zeta_{i}F_{0}
+\xi_{i}^{(1)}(-F_{0} +F_{1}) +2m_{t}^{2}\zeta_{i}G_{0}
  \\ & & -m_{b}^{2}(\xi_{i}^{(1)} +\xi_{i}^{(3)})(G_{1}
-\zeta_{i}G_{0})](m_{b}^{2},m_{i},m_{t})\},
\\ \delta\Lambda_{R}^{(2)} &=& \delta\Lambda_{L}^{(2)}(U),
\hspace{4.0cm} \delta\Lambda_{R}^{(3)} =
\delta\Lambda_{L}^{(3)}(U).
\end{eqnarray*}
Here $C_{0},C_{ij}$ are the three-point Feynman integrals[19] and
$\overline{C}_{24}\equiv-\frac{1}{4}\Delta+C_{24}$, while
\begin{eqnarray*}
F_{n}(q,m_{1},m_{2})&=&\int_{0}^{1}dyy^{n}\ln{[\frac{-q^{2}y(1-y)
+m_{1}^{2}(1-y)+m_{2}^{2}y}{\mu^{2}}]},
\\ G_{n}(q,m_{1},m_{2})
&=&-\int_{0}^{1}dy\frac{y^{n+1}(1-y)} {-q^{2}y(1-y)
+m_{1}^{2}(1-y) +m_{2}^{2}y},
\end{eqnarray*}
and
\begin{eqnarray*}
g_{V}^{t} = \frac{1}{2}-\frac{4}{3}\sin^{2}\theta_{W}, \ \ \ \
g_{A}^{t}=\frac{1}{2}, \ \ \ \ \ \ \  g_{V}^{b} = -\frac{1}{2}
+\frac{2}{3}\sin^{2}\theta_{W}, \ \ \ \ g_{A}^{b} =-\frac{1}{2},
\end{eqnarray*}
which are the SM couplings of the top and bottom quarks to the Z
boson. The definitions of $\theta_q, U_{ij}, V_{ij}, N_{ij}, \mu,
A_{q}$ can be found in ref.[2].

\eject
\begin{center}{\Large Appendix B} \end{center}
\begin{eqnarray*}
h_{1}^{(i)} &=& 4m_{t}\eta^{(2)}(2p_{b}\cdot k -p^{(i)}\cdot p_b)
-4m_{b}\eta^{(1)}(p^{(i)}\cdot p_{t} +p_{t}\cdot k),
\\ h_{2}^{(i)} &=& h_{1}^{(i)}(\eta^{(1)}
\leftrightarrow \eta^{(2)}),
\\ h_{3}^{(i)} &=& 2\eta^{(2)}(2p_{b}\cdot kp_{b}\cdot p_{t}
-m_{b}^{2}p_{t}\cdot k -2p^{(i)}\cdot p_bp_{b}\cdot p_{t})
+2m_{b}m_{t}\eta^{(1)}(p_{b}\cdot k
  \\ & & -2p^{(i)}\cdot p_b),
\\ h_{4}^{(i)} &=& h_{3}^{(i)}(\eta^{(1)}
\leftrightarrow \eta^{(2)}),
\\ h_{5}^{(i)} &=& 2\eta^{(2)}(m_{t}^{2}p_{b}\cdot k
-2p^{(i)}\cdot p_tp_{b}\cdot p_{t})
+2m_{b}m_{t}m_{t}\eta^{(1)}(p_{t}\cdot k -2p^{(i)}\cdot p_t),
\\ h_{6}^{(i)} &=& h_{5}^{(i)}(\eta^{(1)}
\leftrightarrow \eta^{(2)}),
\\ h_{7}^{(i)} &=& 4\eta^{(2)}(p^{(i)}\cdot
p_bp_{t}\cdot k -p^{(i)}\cdot kp_{b}\cdot p_{t} -p_b\cdot
kp^{(i)}\cdot p_t -2p_{b}\cdot kp_{t}\cdot k)
  \\ & & -4m_{b}m_{t}\eta^{(1)} p^{(i)}\cdot k,
\\ h_{8}^{(i)} &=& h_{7}^{(i)}(\eta^{(1)}
\leftrightarrow \eta^{(2)}),
\\ h_{9}^{(i)} &=& 4m_{t}\eta^{(2)} p_{b}\cdot k(p_{b}\cdot k
-p^{(i)}\cdot p_b) -4m_{b}\eta^{(1)}p^{(i)}\cdot p_bp_{t}\cdot k,
\\ h_{10}^{(i)} &=& h_{9}^{(i)}(\eta^{(1)}
\leftrightarrow \eta^{(2)}),
\\ h_{11}^{(i)} &=& 4m_{t}\eta^{(2)} p_{b}\cdot k(p_{t}\cdot k
-p^{(i)}\cdot p_t) -4m_{b}\eta^{(1)} p_{t}\cdot kp^{(i)}\cdot p_t,
\\ h_{12}^{(i)} &=&h_{11}^{(i)}(\eta^{(1)}
\leftrightarrow\eta^{(2)}),
\end{eqnarray*}
where the index $i$ represents the two channels $s$ and $t$, and
$p^{(s)}=p_b$, $p^{(t)}=p_t$. \eject \baselineskip=0.25in {\LARGE
References} \vspace{0.2cm}
\begin{itemize}
\item[{\rm[1]}] For a review, see J.Gunion, H. Haber, G. Kane, and
            S.Dawson, The Higgs Hunter's Guide(Addison-Wesley,
            New York,1990).
\item[{\rm[2]}] H.E. Haber and G.L. Kane, Phys. Rep. 117, 75(1985);
            J.F. Gunion and H.E. Haber, Nucl. Phys. {\bf B272}, 1(1986).
\item[{\rm[3]}] E.Eichten, I.Hinchliffe, K. Lane, and C. Quigg, Rev.
            Mod. Phys. 56, 579(1984); 1065(E)(1986); N.G. Deshpande, X.
Tata,
            and D. A. Dicus, Phys. Rev. {\bf D29}, 1527(1984); S.
Willenbrock,
            Phys. Rev. {\bf D35}, 173(1987); A. Krause, T.Plehn, M. Spria,
            and P. M. Zerwas, Nucl. Phys. {\bf B519}, 85(1998); J.Yi, M.
Wen-Gan,
            H.Liang, H. Meng, and Y. Zeng-Hui, J. Phys. G23, 385(1997);
            Erratum-ibid. G23, 1151(1997).
\item[{\rm[4]}] D.A.Dicus, J.L.Hewett, C.Kao and T.G.Rizzo, Phys. Rev. {\bf
D40}, 787(1989);
            A.A. Barrientos Bendez$\acute{u}$ and B.A. Kniehl, Phys. Rev.
            {\bf D59}, 015009(1999).
\item[{\rm[5]}] S. Moretti and K. Odagiri, Phys. Rev. {\bf
D59},055008(1999).
\item[{\rm[6]}] Z.Kunszt and F. Zwirner, Nucl. Phys. {\bf B385}, 3(1992),
            and references cited therein.
\item[{\rm[7]}] J.F. Gunion, H.E. Haber, F.E. Paige, W.-K. Tung, and
            S. Willenbrock, Nucl. Phys. {\bf B294},621(1987); R.M.
            Barnett, H.E. Haber, and D.E. Soper, ibid. B306,
            697(1988); F.I. Olness and W.-K. Tung, ibid. {\bf B308},
            813(1988).
\item[{\rm[8]}] V. Barger, R.J.N. Phillips, and D.P. Roy, Phys. Lett.
            {\bf B324}, 236(1994).
\item[{\rm[9]}] C.S. Huang and S.H. Zhu, Phys. Rev. {\bf D60},
            075012(1999).
\item[{\rm[10]}] K. Odagiri, hep-ph/9901432; Phys. Lett. {\bf B452},
327(1999).
\item[{\rm[11]}] D.P. Roy, Phys. Lett. {\bf B459}, 607(1999).
\item[{\rm[12]}] Francesca Borzumati, Jean-Loic Kneur, and Nir
            Polonsky, Phys. Rev. {\bf D60}, 115011(1999).
\item[{\rm[13]}] S. Sirlin, Phys. Rev. {\bf D22}, 971 (1980);
            W. J. Marciano and A. Sirlin,{\sl ibid.} {\bf 22}, 2695(1980);
            {\bf 31}, 213(E) (1985);
            A. Sirlin and W.J. Marciano, Nucl. Phys. {\bf B189}, 442(1981);
            K.I. Aoki et.al., Prog. Theor. Phys. Suppl. {\bf 73}, 1(1982).
\item[{\rm[14]}] A. Mendez and A. Pomarol, Phys.Lett.{\bf B279}, 98(1992).
\item[{\rm[15]}] Particle Data Group, C.Caso {\it et al}, Eur.Phys.J.C 3,
1(1998).
\item[{\rm[16]}] J.Gunion, A.Turski, Phys. Rev. {\bf D39}, 2701(1989);
            {\bf D40}, 2333(1990); J.R.Espinosa, M.Quiros, Phys. Lett. {\bf
            B266}, 389(1991); M.Carena, M.Quiros, C.E.M.Wagner, Nucl. Phys.
            {\bf B461}, 407(1996).
\item[{\rm[17]}] H.L. Lai, et al.(CTEQ collaboration), hep-ph/9903282.

\item[{\rm[18}] C.S.Li, R.J.Oakes, and J.M. Yang, Phys. Rev. {\bf D55},
5780(1997).

\item[{\rm[19]}] G.Passarino and M.Veltman, Nucl. Phys. {\bf B160},
151(1979);
            A.Axelrod, {\sl ibid.} {\bf B209}, 349 (1982); M.Clements {\sl
et al.}, Phys. Rev. {\bf D27}, 570 (1983).

\end{itemize}

\newpage
\pagestyle{empty} \topmargin=-1cm \hoffset=-1.5cm \voffset=0.2cm
\textwidth=160mm \textheight=230mm
\date{\today}
\def\baselinestretch{1.5}
\baselineskip=0.3in
%\begin{document}

\vspace{-0.2cm}
\begin{picture}(120,120)(0,0)
\Gluon(5,100)(25,78){-2.5}{3} \ArrowLine(5,56)(25,78)
\ArrowLine(25,78)(75,78) \Vertex(25,78){1} \Vertex(75,78){1}
\ArrowLine(75,78)(95,100) \DashLine(75,78)(95,56){3}
\Text(95,105)[] {$t$} \Text(98,51)[] {$H^-$} \Text(5,105)[]{$g$}
\Text(5,51)[] {$b$} \Text(48,25)[]{$(a)$}
\end{picture}
\vspace{-0.2cm} \hspace{1.0cm}
\begin{picture}(120,120)(0,0)
\Gluon(5,98)(40,98){-2.5}{4} \ArrowLine(5,58)(40,58)
\ArrowLine(40,58)(40,98) \Vertex(40,98){1} \Vertex(40,58){1}
\ArrowLine(40,98)(75,98) \DashLine(40,58)(75,58){3}
\Text(80,101)[] {$t$} \Text(85,55)[] {$H^-$} \Text(0,101)[]{$g$}
\Text(0,55)[] {$b$} \Text(43,35)[]{$(b)$}
\end{picture}
\vspace{-0.2cm}  \hspace{0.9cm}
\begin{picture}(120,120)(0,0)
\Gluon(5,100)(25,78){-2.5}{3} \ArrowLine(5,56)(13.6,65.4)
\Line(13.6,65.4)(25,78) \ArrowLine(25,78)(75,78) \Vertex(25,78){1}
\Vertex(75,78){1} \ArrowLine(75,78)(95,100)
\DashLine(75,78)(95,56){3} \DashCArc(25,78)(17,-132,0){3}
\Text(95,105)[] {$t$} \Vertex(13.6,65.4){1} \Vertex(42,78){1}
\Text(98,51)[] {$H^-$} \Text(5,105)[]{$g$} \Text(5,51)[] {$b$}
\Text(48,35)[]{$(c)$}
\end{picture}\\
\vspace{-0.2cm}  \hspace{0.4cm}
\begin{picture}(120,120)(0,0)
\Gluon(5,100)(25,78){-2.5}{3} \ArrowLine(5,56)(13.6,65.4)
\DashLine(13.6,65.4)(25,78){3} \DashLine(25,78)(42,78){3}
\ArrowLine(42,78)(75,78) \Vertex(25,78){1} \Vertex(75,78){1}
\ArrowLine(75,78)(95,100) \DashLine(75,78)(95,56){3}
\CArc(25,78)(17,-132,0) \Text(95,105)[] {$t$}
\Vertex(13.6,65.4){1} \Vertex(42,78){1} \Text(98,51)[] {$H^-$}
\Text(5,105)[]{$g$} \Text(5,51)[] {$b$} \Text(48,35)[]{$(d)$}
\end{picture}
\vspace{-0.2cm} \hspace{0.9cm}
\begin{picture}(120,120)(0,0)
\Gluon(5,100)(20,83){-2.5}{3} \ArrowLine(5,56)(20,73)
\ArrowLine(30,78)(75,78) \Line(22,82)(28,74) \Line(22,74)(28,82)
\Vertex(75,78){1} \ArrowLine(75,78)(95,100)
\DashLine(75,78)(95,56){3} \Text(95,105)[] {$t$} \Text(98,51)[]
{$H^-$} \Text(5,105)[]{$g$} \Text(5,51)[] {$b$}
\Text(48,25)[]{$(e)$}
\end{picture}
\vspace{-0.2cm} \hspace{1.1cm}
\begin{picture}(120,120)(0,0)
\Gluon(5,98)(40,98){-2.5}{4} \ArrowLine(5,58)(40,58)
\Line(40,98)(57.5,98) \DashLine(57.5,98)(40,78){3}
\Line(40,98)(40,88) \ArrowLine(40,58)(40,88) \Vertex(40,98){1}
\Vertex(40,58){1} \ArrowLine(57.5,98)(75,98)
\DashLine(40,58)(75,58){3} \Vertex(57.5,98){1} \Vertex(40,78){1}
\Text(80,101)[] {$t$} \Text(85,55)[] {$H^-$} \Text(0,101)[]{$g$}
\Text(0,55)[] {$b$} \Text(43,35)[]{$(f)$}
\end{picture}\\
\vspace{-0.2cm} \hspace{0.4cm}
\begin{picture}(120,120)(0,0)
\Gluon(5,98)(40,98){-2.5}{4} \ArrowLine(5,58)(40,58)
\DashLine(40,98)(57.5,98){3} \Line(57.5,98)(40,78)
\DashLine(40,78)(40,98){3} \ArrowLine(40,58)(40,78)
\Vertex(40,98){1} \Vertex(40,58){1} \ArrowLine(57.5,98)(75,98)
\DashLine(40,58)(75,58){3} \Vertex(57.5,98){1} \Vertex(40,78){1}
\Text(80,101)[] {$t$} \Text(85,55)[] {$H^-$} \Text(0,101)[]{$g$}
\Text(0,55)[] {$b$} \Text(43,35)[]{$(g)$}
\end{picture}
\vspace{-0.2cm} \hspace{1.1cm}
\begin{picture}(120,120)(0,0)
\Gluon(5,98)(35,98){-2.5}{4} \ArrowLine(5,58)(40,58)
\ArrowLine(40,58)(40,93) \Vertex(40,58){1}
\ArrowLine(45,98)(75,98) \DashLine(40,58)(75,58){3}
\Line(37,102)(43,94) \Line(37,94)(43,102) \Text(80,101)[] {$t$}
\Text(85,55)[] {$H^-$} \Text(0,101)[]{$g$} \Text(0,55)[] {$b$}
\Text(43,35)[]{$(h)$}
\end{picture}
\vspace{-0.2cm} \hspace{0.8cm}
\begin{picture}(120,120)(0,0)
\Gluon(5,100)(25,78){-2.5}{3} \ArrowLine(5,56)(25,78)
\DashCArc(50,78)(17,0,180){3} \Line(25,78)(75,78)
\Vertex(25,78){1} \Vertex(75,78){1} \ArrowLine(75,78)(95,100)
\Vertex(33,78){1} \Vertex(67,78){1} \DashLine(75,78)(95,56){3}
\Text(95,105)[] {$t$} \Text(98,51)[] {$H^-$} \Text(5,105)[]{$g$}
\Text(5,51)[] {$b$} \Text(48,35)[]{$(i)$}
\end{picture}\\
\vspace{-0.2cm} \hspace{0.4cm}
\begin{picture}(120,120)(0,0)
\Gluon(5,100)(25,78){-2.5}{3} \ArrowLine(5,56)(25,78)
\Line(25,78)(45,78) \Line(55,78)(75,78) \Line(47,82)(53,74)
\Line(53,82)(47,74) \Vertex(25,78){1} \Vertex(75,78){1}
\ArrowLine(75,78)(95,100) \DashLine(75,78)(95,56){3}
\Text(95,105)[] {$t$} \Text(98,51)[] {$H^-$} \Text(5,105)[]{$g$}
\Text(5,51)[] {$b$} \Text(48,25)[]{$(j)$}
\end{picture}
\vspace{-0.3cm} \hspace{1.1cm}
\begin{picture}(120,120)(0,0)
\Gluon(5,98)(40,98){-2.5}{4} \ArrowLine(5,58)(40,58)
\Line(40,58)(40,98) \Vertex(40,98){1} \Vertex(40,58){1}
\DashCArc(40,78)(13,-90,90){3} \ArrowLine(40,98)(75,98)
\Vertex(40,91){1} \Vertex(40,65){1} \DashLine(40,58)(75,58){3}
\Text(80,101)[] {$t$} \Text(85,55)[] {$H^-$} \Text(0,101)[]{$g$}
\Text(0,55)[] {$b$} \Text(43,35)[]{$(k)$}
\end{picture}
\vspace{-0.3cm} \hspace{1.0cm}
\begin{picture}(120,120)(0,0)
\Gluon(5,98)(40,98){-2.5}{4} \ArrowLine(5,58)(40,58)
\Line(40,58)(40,73) \Line(40,83)(40,98) \Vertex(40,98){1}
\Vertex(40,58){1} \ArrowLine(40,98)(75,98) \Line(37,82)(43,74)
\Line(37,74)(43,82) \DashLine(40,58)(75,58){3} \Text(80,101)[]
{$t$} \Text(85,55)[] {$H^-$} \Text(0,101)[]{$g$} \Text(0,55)[]
{$b$} \Text(43,35)[]{$(l)$}
\end{picture}\\
\vspace{-0.3cm} \hspace{0.4cm}
\begin{picture}(120,120)(0,0)
\Gluon(5,100)(25,78){-2.5}{3} \ArrowLine(5,56)(25,78)
\ArrowLine(25,78)(75,78) \Vertex(25,78){1} \Vertex(75,78){1}
\DashCArc(75,78)(17,48,180){3} \ArrowLine(86.4,90.6)(95,100)
\Line(75,78)(86.4,90.6) \Vertex(58,78){1} \Vertex(86.4,90.6){1}
\DashLine(75,78)(95,56){3} \Text(95,105)[] {$t$} \Text(98,51)[]
{$H^-$} \Text(5,105)[]{$g$} \Text(5,51)[] {$b$}
\Text(48,35)[]{$(m)$}
\end{picture}
\vspace{-0.3cm} \hspace{1.0cm}
\begin{picture}(120,120)(0,0)
\Gluon(5,100)(25,78){-2.5}{3} \ArrowLine(5,56)(25,78)
\ArrowLine(25,78)(75,78) \Vertex(25,78){1} \Vertex(75,78){1}
\ArrowLine(75,78)(95,100) \DashLine(75,78)(95,56){3}
\DashCArc(75,78)(17,-180,-48){3} \Vertex(86.4,65.4){1}
\Vertex(58,78){1} \Text(95,105)[] {$t$} \Text(98,51)[] {$H^-$}
\Text(5,105)[]{$g$} \Text(5,51)[] {$b$} \Text(92,74)[]{\small
$\tilde{t}_j$} \Text(72,55)[]{\small $\tilde{b}_i$}
\Text(65,87)[]{\small $\tilde\chi_k^0$} \Text(48,35)[]{$(n)$}
\end{picture}
\vspace{-0.3cm} \hspace{1.0cm}
\begin{picture}(120,120)(0,0)
\Gluon(5,100)(25,78){-2.5}{3} \ArrowLine(5,56)(25,78)
\ArrowLine(25,78)(70,78) \Vertex(25,78){1}
\ArrowLine(80,82)(95,100) \DashLine(80,74)(95,56){3}
\Line(72,82)(78,74) \Line(72,74)(78,82) \Text(95,105)[] {$t$}
\Text(98,51)[] {$H^-$} \Text(5,105)[]{$g$} \Text(5,51)[] {$b$}
\Text(48,25)[]{$(o)$}
\end{picture}\\
\vspace{-0.3cm} \hspace{0.4cm}
\begin{picture}(120,120)(0,0)
\Gluon(5,98)(40,98){-2.5}{4} \ArrowLine(5,58)(22.5,58)
\Line(22.5,58)(40,58) \Line(40,58)(40,68) \ArrowLine(40,68)(40,98)
\Vertex(40,98){1} \DashLine(22.5,58)(40,78){3} \Vertex(40,58){1}
\Vertex(40,78){1} \Vertex(22.5,58){1} \ArrowLine(40,98)(75,98)
\DashLine(40,58)(75,58){3} \Text(80,101)[] {$t$} \Text(85,55)[]
{$H^-$} \Text(0,101)[]{$g$} \Text(0,55)[] {$b$}
\Text(43,35)[]{$(p)$}
\end{picture}
\vspace{-0.3cm}\hspace{1.1cm}
\begin{picture}(120,120)(0,0)
\Gluon(5,98)(40,98){-2.5}{4} \ArrowLine(5,58)(22.5,58)
\ArrowLine(40,78)(40,98) \Vertex(40,98){1} \Vertex(40,58){1}
\DashLine(40,58)(40,78){3} \Line(22.5,58)(40,78)
\ArrowLine(40,98)(75,98) \DashLine(22.5,58)(75,58){3}
\Vertex(40,78){1} \Vertex(22.5,58){1} \Text(80,101)[] {$t$}
\Text(85,55)[] {$H^-$} \Text(0,101)[]{$g$} \Text(0,55)[] {$b$}
\Text(34,50)[]{\small $\tilde{b}_i$} \Text(48,68)[]{\small
$\tilde{t}_j$} \Text(23,72)[]{\small $\tilde\chi_k^0$}
\Text(43,35)[]{$(q)$}
\end{picture}
\vspace{-0.3cm}\hspace{1.1cm}
\begin{picture}(120,120)(0,0)
\Gluon(5,98)(40,98){-2.5}{4} \ArrowLine(5,58)(35,58)
\ArrowLine(40,63)(40,98) \Vertex(40,98){1} \Line(37,62)(43,54)
\Line(37,54)(43,62) \ArrowLine(40,98)(75,98)
\DashLine(45,58)(75,58){3} \Text(80,101)[] {$t$} \Text(85,55)[]
{$H^-$} \Text(0,101)[]{$g$} \Text(0,55)[] {$b$}
\Text(43,35)[]{$(r)$}
\end{picture}\\
\vspace{-0.3cm} \hspace{0.4cm}
\begin{picture}(120,120)(0,0)
\Gluon(5,98)(32,98){-2.5}{3} \ArrowLine(5,58)(32,58)
\DashLine(32,58)(32,98){3} \Vertex(32,98){1} \Vertex(32,58){1}
\DashLine(32,98)(58,98){3} \DashLine(58,98)(85,98){3}
\Line(32,58)(58,58) \Line(58,98)(58,58) \DashLine(58,58)(85,58){3}
\Vertex(58,98){1} \Vertex(58,58){1} \Text(90,55)[] {$t$}
\Text(95,101)[] {$H^-$} \Text(0,101)[]{$g$} \Text(0,55)[] {$b$}
\Text(25,78)[]{\small $\tilde{b}_l$} \Text(65,78)[]{\small
$\tilde{t}_j$} \Text(45,110)[]{\small $\tilde{b}_i$}
\Text(45,50)[]{\small $\tilde\chi_k^0$} \Text(48,35)[]{$(s)$}
\end{picture}
\vspace{-0.3cm} \hspace{1.1cm}
\begin{picture}(120,120)(0,0)
\Gluon(5,98)(32,98){-2.5}{3} \ArrowLine(5,58)(32,58)
\ArrowLine(32,58)(32,98) \Vertex(32,98){1} \Vertex(32,58){1}
\ArrowLine(32,98)(58,98) \DashLine(58,98)(85,98){3}
\DashLine(32,58)(58,58){3} \ArrowLine(58,98)(58,58)
\ArrowLine(58,58)(85,58) \Vertex(58,98){1} \Vertex(58,58){1}
\Text(90,55)[] {$t$} \Text(95,101)[] {$H^-$} \Text(0,101)[]{$g$}
\Text(0,55)[] {$b$} \Text(25,78)[]{$b$} \Text(65,78)[]{$t$}
\Text(48,35)[]{$(t)$}
\end{picture}
\vspace{-0.3cm}\hspace{0.8cm}
\begin{picture}(120,120)(0,0)
\Gluon(5,98)(32,98){-2.5}{3} \DashLine(5,58)(32,58){3}
\ArrowLine(32,58)(32,98) \Vertex(32,98){1} \Vertex(32,58){1}
\ArrowLine(32,98)(58,98) \ArrowLine(58,98)(85,98)
\ArrowLine(58,58)(32,58) \ArrowLine(85,58)(58,58)
\DashLine(58,58)(58,98){3} \Vertex(58,98){1} \Vertex(58,58){1}
\Text(90,101)[] {$t$} \Text(90,55)[] {$b$} \Text(0,101)[]{$g$}
\Text(25,78)[]{$t$} \Text(45,50)[]{$b$} \Text(0,55)[] {$H^-$}
\Text(48,35)[]{$(u)$}
\end{picture}\\
\vspace{-0.3cm}\hspace{0.6cm}
\begin{picture}(120,120)(0,0)
\Gluon(5,98)(32,98){-2.5}{3} \DashLine(5,58)(32,58){3}
\DashLine(32,58)(32,98){3} \Vertex(32,98){1} \Vertex(32,58){1}
\DashLine(32,98)(58,98){3} \ArrowLine(58,98)(85,98)
\DashLine(58,58)(32,58){3} \ArrowLine(85,58)(58,58)
\Line(58,58)(58,98) \Vertex(58,98){1} \Vertex(58,58){1}
\Text(90,101)[] {$t$} \Text(90,55)[] {$b$} \Text(0,101)[]{$g$}
\Text(25,78)[]{\small $\tilde{t}_j$} \Text(45,50)[]{\small
$\tilde{b}_i$} \Text(0,55)[] {$H^-$} \Text(68,78)[]{\small
$\tilde{t}_l$} \Text(45,107)[]{\small
$\tilde\chi_k^0$}\Text(48,35)[]{$(v)$}
\end{picture}

{\small \begin{figure}[ht]
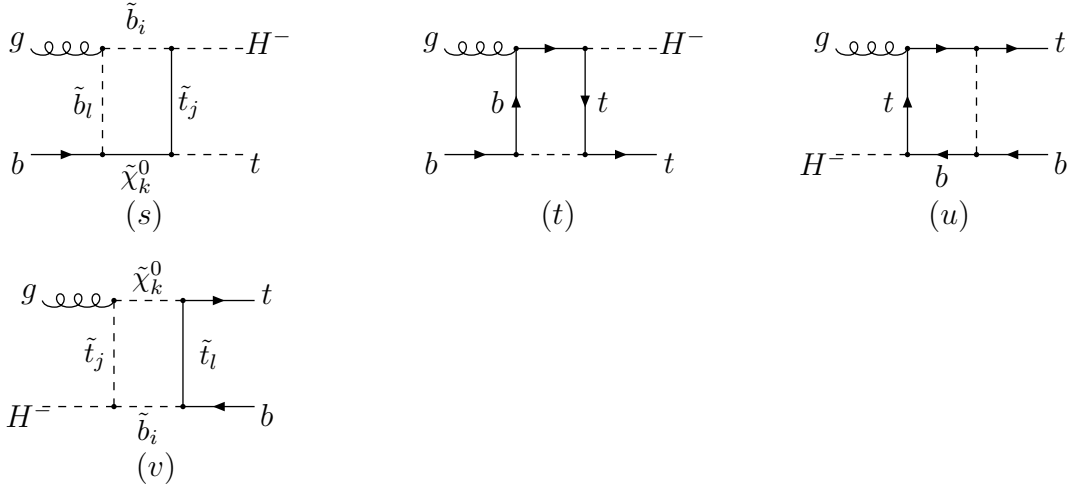
 \caption[]{ \small Feynman diagrams
contributing to $O(\alpha_{ew} m_{t(b)}^{2}/m_{W}^{2})$ Yukawa
corrections to $gb\rightarrow tH^{-}$: $(a)$ and $(b)$ are tree level
diagrams; $(c)-(v)$ are one-loop diagrams. The dashed lines
represent $H,h,A,H^{\pm},G^{0}$ and $G^{\pm}$ for diagrams $(c)$ and $(f)$;
$H,h,A$ and $G^{0}$ for diagrams $(m),(p),(t)$ and $(u)$; 
$\tilde{t},\tilde{b},H,h,A,H^{\pm},G^{0}$ and $G^{\pm}$ for 
$(i)$ and $(k)$, 
where the solid lines represent charginos and
neutralinos if the dashed lines represent squarks.
For diagrams $(d)$ and $(g)$, the solid lines in the loop represent
$\tilde\chi^0$ and $\tilde\chi^+$
 and the dashed lines represent squarks.}
\end{figure}}

\newpage
\vspace{0.1cm} \hspace{0.1cm}
\begin{picture}(120,120)(0,0)
\ArrowLine(0,62)(25,62) \Line(25,62)(75,62)
\ArrowLine(75,62)(100,62) \DashCArc(50,62)(25,0,180){3}
\Vertex(25,62){1} \Vertex(75,62){1} \Text(5,53)[] {$t(b)$}
\Text(103,53)[] {$t(b)$} \Text(53,5)[]{$(a)$}
\end{picture}
\vspace{0.1cm}  \hspace{0.8cm}
\begin{picture}(120,120)(0,0)
\Photon(5,60)(33,60){2.5}{4} \Photon(77,60)(105,60){2.5}{4}
\ArrowArc(55,60)(22,0,180) \ArrowArc(55,60)(22,180,360)
\Vertex(33,60){1} \Vertex(77,60){1} \Text(5,50)[]{$W^{-}$}
\Text(108,50)[]{$W^{-}$} \Text(55,93)[]{$t$} \Text(55,28)[]{$b$}
\Text(55,5)[]{$(b)$}
\end{picture}
\vspace{0.1cm}  \hspace{0.8cm}
\begin{picture}(120,120)(0,0)
\Photon(5,60)(33,60){2.5}{4} \Photon(77,60)(105,60){2.5}{4}
\ArrowArc(55,60)(22,0,180) \ArrowArc(55,60)(22,180,360)
\Vertex(33,60){1} \Vertex(77,60){1} \Text(5,50)[]{$Z^{0}$}
\Text(108,50)[]{$Z^{0}$} \Text(55,93)[]{$t(b)$}
\Text(55,28)[]{$t(b)$} \Text(55,5)[]{$(c)$}
\end{picture}\\
\vspace{-0.5cm} \hspace{0.5cm}
\begin{picture}(120,120)(0,0)
\DashLine(5,80)(33,80){3} \Photon(77,80)(105,80){2.5}{4}
\ArrowArc(55,80)(22,0,180) \ArrowArc(55,80)(22,180,360)
\Vertex(33,80){1} \Vertex(77,80){1} \Text(5,70)[]{$H^{-}$}
\Text(108,70)[]{$W^{-}$} \Text(55,113)[]{$t$} \Text(55,48)[]{$b$}
\Text(55,25)[]{$(d)$}
\end{picture}
\vspace{-0.5cm} \hspace{1.0cm}
\begin{picture}(120,120)(0,0)
\DashLine(5,80)(33,80){3} \Photon(77,80)(105,80){2.5}{4}
\DashCArc(55,80)(22,0,360){3} \Vertex(33,80){1} \Vertex(77,80){1}
\Text(5,70)[]{$H^{-}$} \Text(108,70)[]{$W^{-}$}
\Text(55,113)[]{\small $\tilde{t}_j$} \Text(55,48)[]{\small
$\tilde{b}_i$} \Text(55,25)[]{$(e)$}
\end{picture}
\vspace{-0.5cm}  \hspace{1.0cm}
\begin{picture}(120,120)(0,0)
\DashLine(5,80)(33,80){3} \DashLine(77,80)(105,80){3}
\ArrowArc(55,80)(22,0,180) \ArrowArc(55,80)(22,180,360)
\Vertex(33,80){1} \Vertex(77,80){1} \Text(5,70)[]{$H^{-}$}
\Text(108,70)[]{$H^{-}$} \Text(55,113)[]{$t$} \Text(55,48)[]{$b$}
\Text(55,25)[]{$(f)$}
\end{picture}\\
% new

\vskip 1cm
\vspace{-0.5cm}\hspace{0.1cm}
\begin{picture}(120,120)(0,0)
\DashLine(5,80)(33,80){3} \DashLine(77,80)(105,80){3}
\DashCArc(55,80)(22,0,360){3} \Vertex(33,80){1} \Vertex(77,80){1}
\Text(5,70)[]{$H^{-}$} \Text(108,70)[]{$H^{-}$}
\Text(55,113)[]{\small $\tilde{t}_j$} \Text(55,48)[]{\small
$\tilde{b}_i$} \Text(55,25)[]{$(g)$}
\end{picture}

{\small \begin{figure}[ht] \caption[]{ \small Self-energy
Feynman diagrams
contributing to renormalization constants: The dashed lines
represent $\tilde{t},\tilde{b},H,h,A,H^{\pm},G^{0}$ and $G^{\pm}$ for
diagram $(a)$,
where the solid lines represent charginos and
neutralinos if the dashed lines represent squarks. }
\end{figure}}

\begin{figure}
\epsfxsize=15 cm
\centerline{
\epsffile{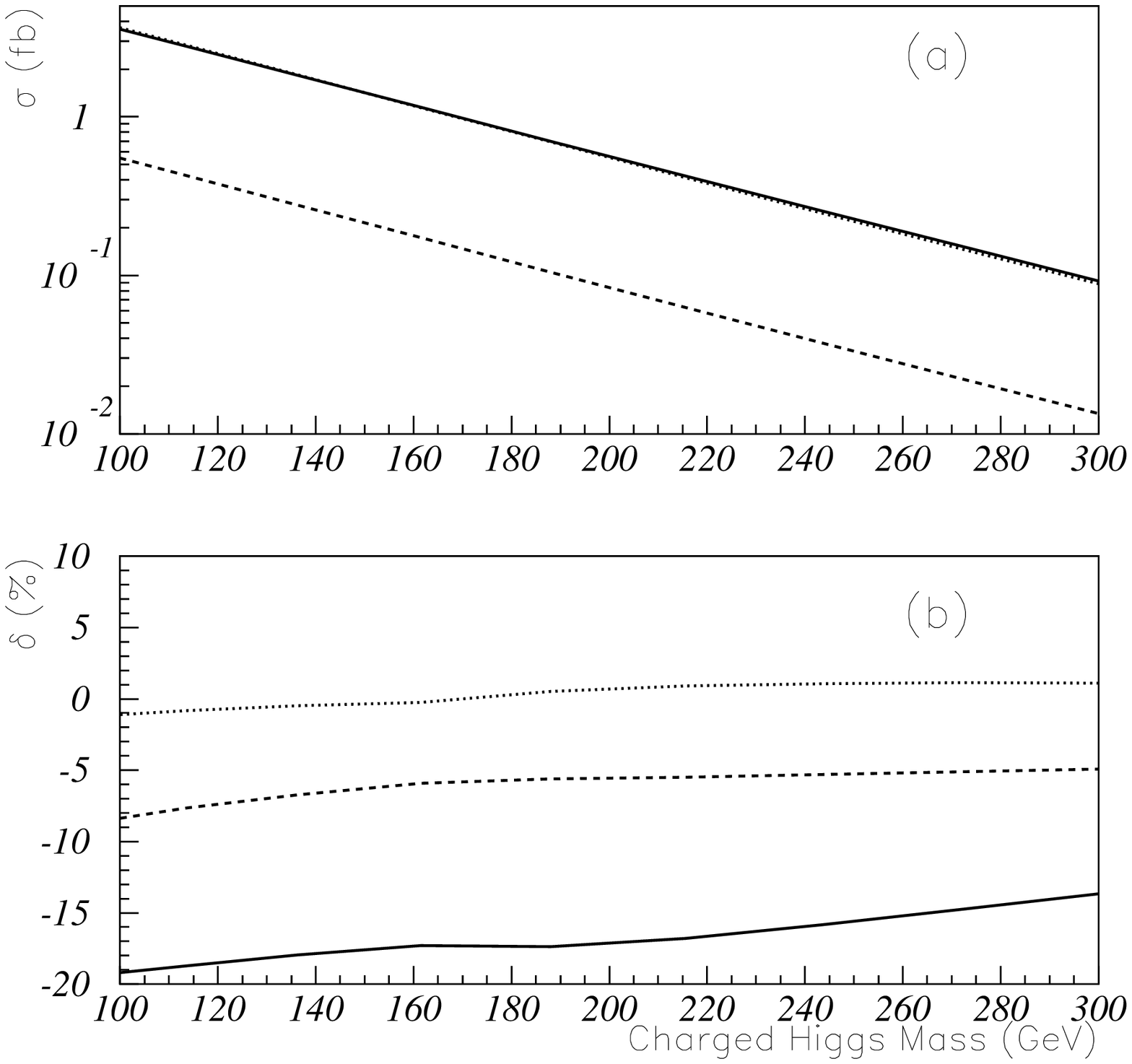
}}
\caption[]{
    The tree-level total cross sections (a) and relative one-loop 
    corrections (b) versus $m_{H^{\pm}}$ at the Tevatron with 
    $\sqrt{s}= 2$  TeV. The solid, dashed and dotted lines correspond to 
    $\tan\beta=2,10$ and $30$, respectively.}
\end{figure}
 
\begin{figure}
\epsfxsize=15 cm
\centerline{
\epsffile{
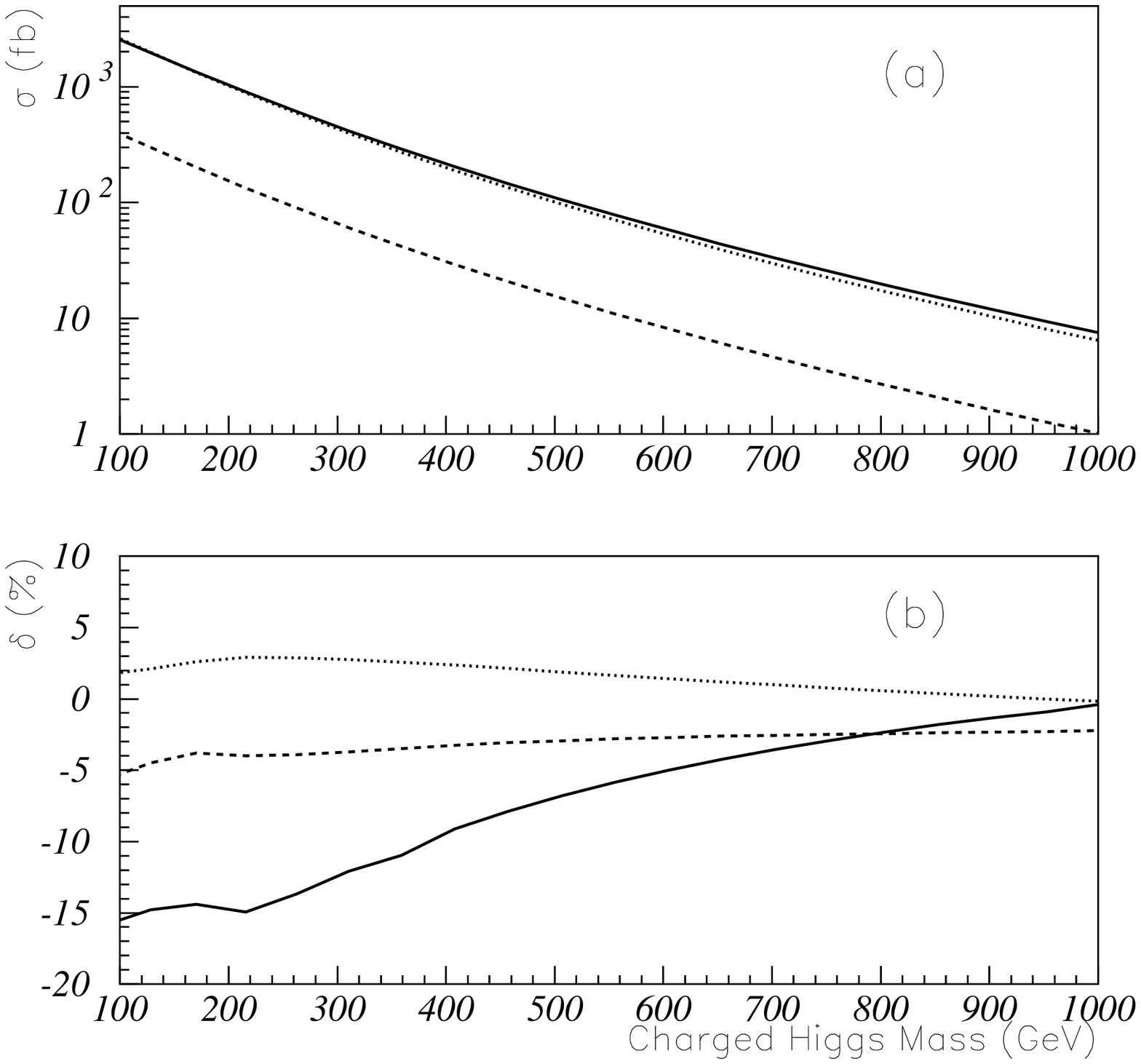
}
}
\caption[]{
    The tree-level total cross sections (a) and relative one-loop
    corrections (b) versus $m_{H^{\pm}}$ at the LHC with 
    $\sqrt{s}= 14$  TeV. The solid, dashed and dotted lines correspond to 
    $\tan\beta=2,10$ and $30$, respectively.}
\end{figure}
 
\begin{figure}
\epsfxsize=15 cm
\centerline{
\epsffile{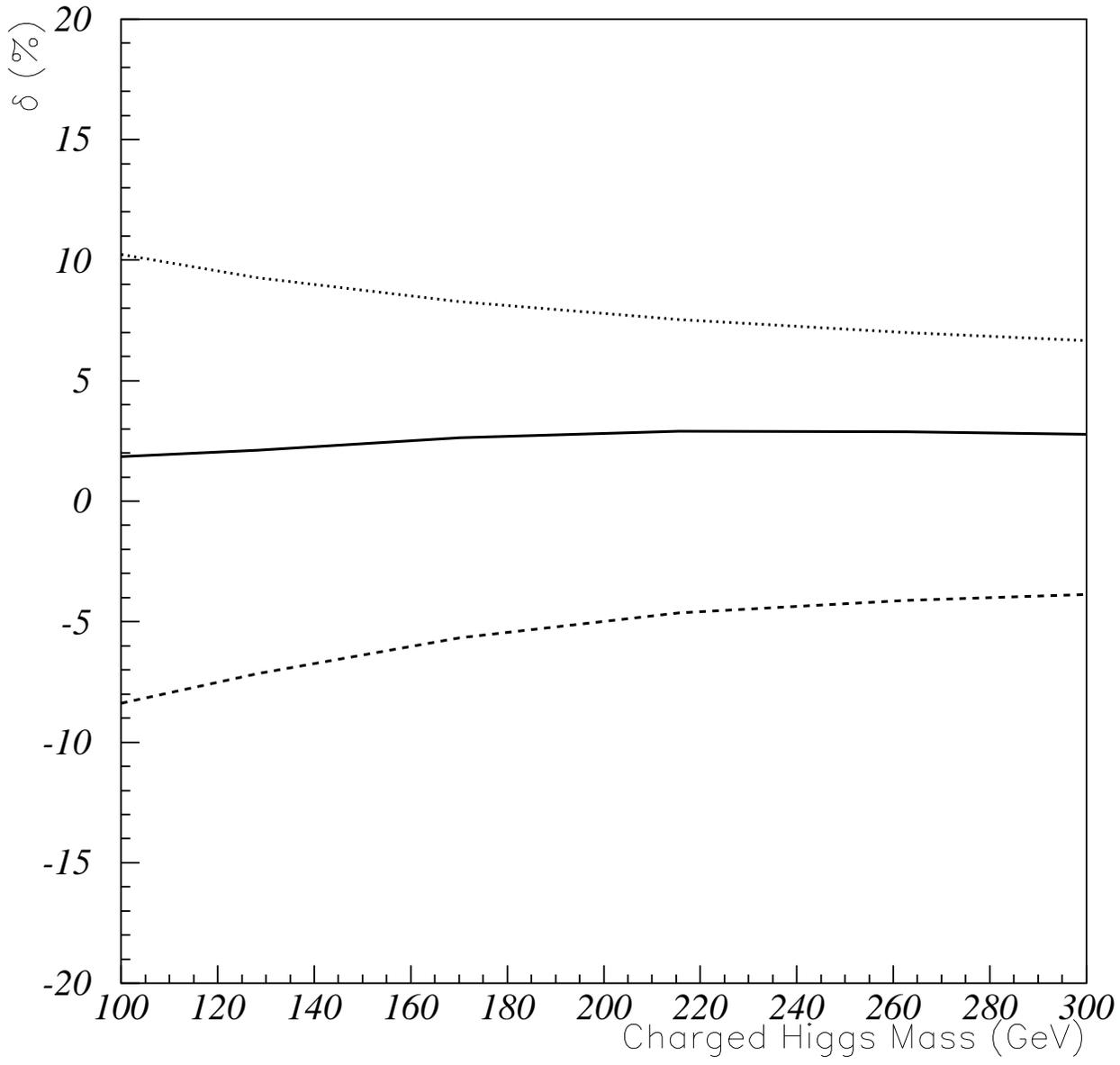
}}
\caption[]{
The radiative correction from top, bottom quarks (dashed line)
 and genuine SUSY
particles (dotted line), as well as total contributions (solid line)
 when  $\tan\beta=30$
 at the LHC with
    $\sqrt{s}= 14$  TeV.}
\end{figure}
 
\begin{figure}
\epsfxsize=15 cm
\centerline{
\epsffile{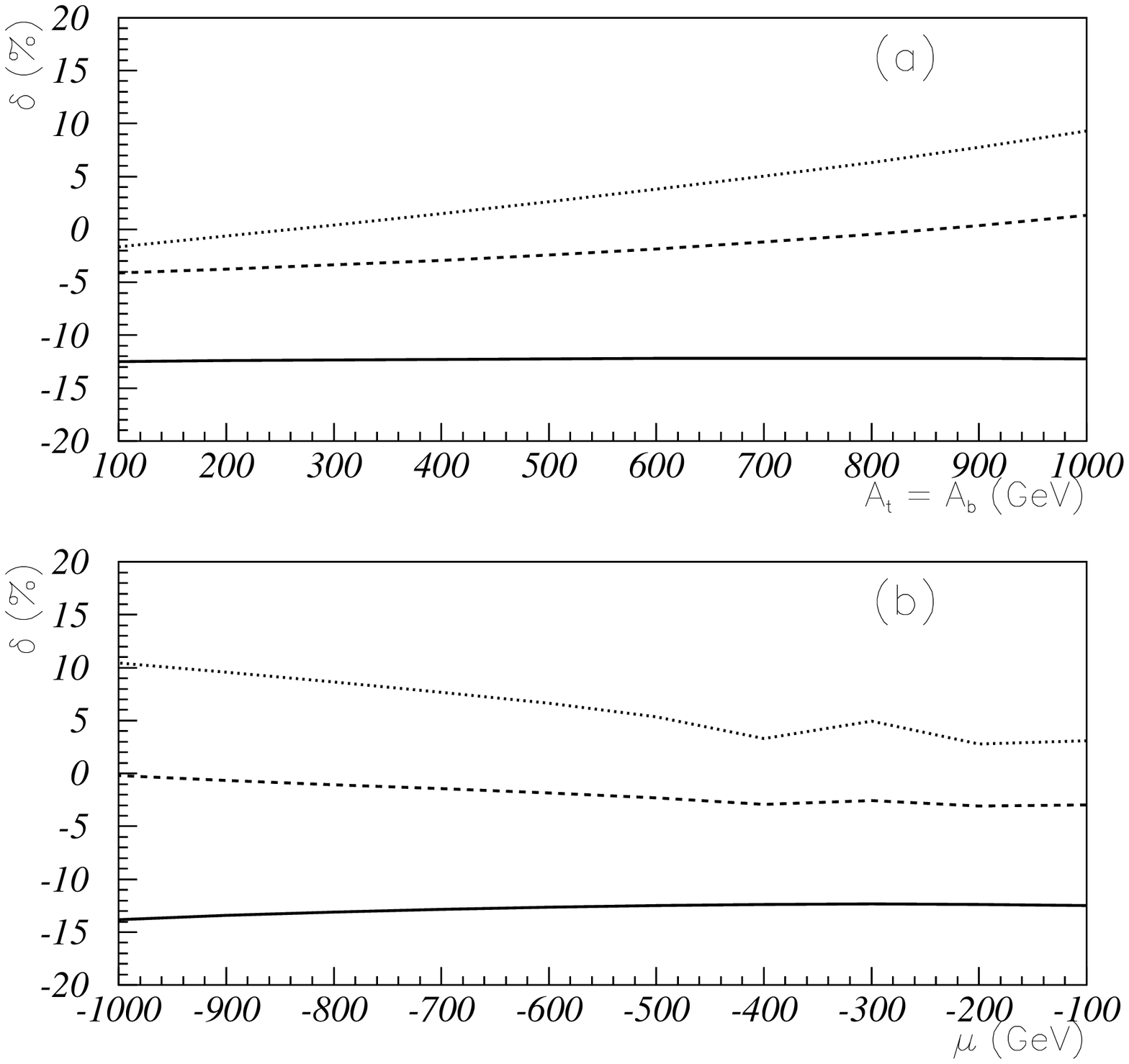
}}
\caption[]{
    Relative one-loop 
    corrections versus $A_t$, $A_b$ (a) as well as
$\mu$ (b) at the LHC with
    $\sqrt{s}= 14$  TeV,
where $m_{H^{\pm}}=300 GeV$ and the solid, 
dashed and dotted lines correspond to 
    $\tan\beta=2,10$ and $30$, respectively.
For (a), $\mu=-100 GeV$, and for (b), $A_t=A_b=200 GeV$.}
\end{figure}
 
%\end{document} 

\end{document}